\documentclass[amsmath, amssymb, twocolumn, 10pt, prb]{revtex4-1}

\usepackage{appendix}

\usepackage{amssymb, amsmath,amsthm}    
\usepackage{graphicx}   
\usepackage{verbatim}   
\usepackage{color}      
\usepackage{subfigure}  
\usepackage{graphicx}
\usepackage{caption}
\usepackage{tikz}
\usepackage{pgflibraryarrows}
\usepackage{pgflibrarysnakes}
\usepackage{mathtools}
\usetikzlibrary{decorations.markings}
\usetikzlibrary{decorations.footprints}
\definecolor{green}{RGB}{35,142,35}
\usetikzlibrary{decorations.pathmorphing}
\usepackage{xcolor}

\usepackage{ae} 
\usepackage[T1]{fontenc}
\usepackage[ansinew]{inputenc}
\usepackage{amsmath}
\usepackage{amsthm}
\usepackage{amssymb}
\usepackage{graphicx}
\usepackage{color}
\definecolor{darkblue}{cmyk}{0.9,0.9,0,0}
\usepackage[pdfstartview=FitV,linktocpage,colorlinks=true,linkcolor=darkblue,citecolor=darkblue,urlcolor=darkblue]{hyperref}
\usepackage{epsfig}
\usepackage{enumerate}
\usepackage{tikz}
\usepackage{subfigure}
\usepackage{verbatim}
\usepackage{cleveref}

\usepackage{varioref}
\usepackage{makeidx}
\makeindex

\usepackage[english]{babel}

\newcommand{\ncmd}{\newcommand}
\ncmd{\lt}{\left}
\ncmd{\rt}{\right}

\newcommand{\beq}{\begin{equation}}
\newcommand{\eeq}{\end{equation}}
\newcommand{\beqq}{\begin{equation*}}
\newcommand{\eeqq}{\end{equation*}}
\newcommand\beqa{\begin{eqnarray}}
\newcommand\eeqa{\end{eqnarray}}
\newcommand\beqaa{\begin{eqnarray*}}
\newcommand\eeqaa{\end{eqnarray*}}
\newcommand\bea{\begin{array}}
\newcommand\eea{\end{array}}
\newcommand\beaa{\begin{array*}}
\newcommand\eeaa{\end{array*}}
\newcommand\beal{\begin{align}}
\newcommand\eeal{\end{align}}
\newcommand\beall{\begin{align*}}
\newcommand\eeall{\end{align*}}

\newcommand{\eqaln}[1]{\begin{align}#1\end{align}}

\def\XXint#1#2#3{{\setbox0=\hbox{$#1{#2#3}{\int}$}
\vcenter{\hbox{$#2#3$}}\kern-.5\wd0}}

\newcommand{\ep}{\epsilon}
\newcommand{\eps}{\varepsilon}
\newcommand{\fr}{\frac}
\newcommand{\f}{\frac}

\newcommand{\nn}{\nonumber}

\newcommand{\neqa}{\nonumber\end{eqnarray}}

\newcommand{\eq}[1]{Eq.(\ref{#1})}

\newcommand{\Tr}{{\rm Tr}}

\newcommand{\half}{\frac{1}{2}}

\renewcommand{\d}{\partial}

\newcommand{\<}{{\langle}}
\renewcommand{\>}{{\rangle}}

\newcommand{\abs}[1]{\lt|{#1}\rt|}

\newcommand{\tr}[1]{\mbox{Tr}\lt[{#1}\rt]}
\newcommand{\mc}{\mathcal}

\newcommand{\re}{\relax{\rm I\kern-.18em R}}

\renewcommand{\sp}{p\hspace{-.40em}/}

\def\su2{{SU(2)}}

\makeatletter
\newcommand*{\rom}[1]{\expandafter\@slowromancap\romannumeral #1@}
\makeatother

\def\a{{\alpha}}

\def\[{\left[}
\def\]{\right]}
\def\({\left(}
\def\){\right)}

\def\l{\lambda}
\def\k{\kappa}

\def\s{\sigma}
\def\a{\alpha}

\def\G{\Gamma}
\def\g{\gamma}

\def\({\left(}
\def\){\right)}
\def\[{\left[}
\def\]{\right]}

\def\<{\langle}
\def\>{\rangle}

\def\i2{\frac{i}{2}}

\def\spi{\relax{\rm \pi\kern-0.5em /}}
\def\sA{\relax{\rm A\kern-0.5em /}}
\def\sp{\relax{\rm p\kern-0.5em /}}
\def\sd{\relax{\rm \d\kern-0.5em /}}
\def\sk{\relax{\rm k\kern-0.5em /}}
\def\sn{\relax{\rm n\kern-0.5em /}}
\def\sl{\relax{\rm l\kern-0.5em /}}
\def\sP{\relax{\rm P\kern-0.7em /}}
\def\sBethe{\relax{\rm \Bethe\kern-0.5em /}}

\renewcommand\theequation{{\color{blue}\arabic{equation}}}
\newcommand{\dd}{\mathrm{d}}

\begin{document}
	\title{Emergence of a control parameter for the antiferromagnetic quantum critical metal}
	\author{Peter Lunts}
	\author{Andres Schlief}
	\author{Sung-Sik Lee}
	\affiliation{Perimeter Institute for Theoretical Physics, 31 Caroline St. N., Waterloo ON N2L 2Y5, Canada}
	\affiliation{Department of Physics \& Astronomy, McMaster University, 1280 Main St. W., Hamilton ON L8S 4M1, Canada}
	\date{\today}

\begin{abstract}
We study the antiferromagnetic quantum critical metal 
in $3-\epsilon$ space dimensions by extending
the earlier one-loop analysis
[Sur and Lee, Phys. Rev. B {\bf 91}, 125136 (2015)]
 to higher-loop orders.
We show that the $\ep$-expansion is not organized by
the standard loop expansion,
and a two-loop graph becomes as important as one-loop graphs
due to an infrared singularity caused by an emergent quasi-locality.
This qualitatively changes the nature of the infrared (IR) fixed point,
and the $\ep$-expansion is controlled only after 
the two-loop effect is taken into account.
Furthermore, we show that a ratio between velocities emerges as a small parameter, 
which suppresses a large class of diagrams.
We show that the critical exponents do not receive corrections beyond the linear order in $\ep$ 
in the limit that the ratio of velocities vanishes.
The $\ep$-expansion gives critical exponents 
which are consistent with the exact solution
obtained in $0 < \ep \leq 1$.
\end{abstract}
	
	\maketitle

\section{Introduction}

Quantum critical points in metals 
host unconventional metallic states
which lie outside the realm 
of Landau Fermi liquid theory\cite{LFL, RevModPhys.66.129, 1992hep.th...10046P}.
Experimentally, 
non-Fermi liquids 
are often characterized by
anomalous dependencies 
of thermodynamic, spectroscopic and transport properties
on temperature and energy
\cite{
RevModPhys.79.1015, 
RevModPhys.73.797}.
On the theoretical side,
the quasiparticle paradigm
based on well-defined 
single-particle excitations
needs to be replaced with 
theories that capture 
strong interactions
between soft collective modes
and electronic excitations
\cite{
PhysRevB.8.2649,
PhysRevB.14.1165, 
PhysRevB.40.11571,
PhysRevLett.63.680,
PhysRevB.46.5621,
PhysRevB.48.7183, 
PhysRevB.50.14048,
PhysRevB.50.17917,
polchinski1994low,
nayak1994non,
PhysRevB.78.035103,
PhysRevB.80.165102,
PhysRevB.82.075127,
PhysRevB.82.045121,
jiang2013non,
PhysRevB.88.125116,
PhysRevB.88.245106,
PhysRevB.90.045121,
PhysRevB.92.041112,
PhysRevX.6.031028}.

The antiferromagnetic (AF) quantum phase transition arises 
in a wide range of strongly correlated materials such as 
electron doped cuprates\cite{PhysRevLett.105.247002}, iron pnictides\cite{Hashimoto22062012}, and heavy fermion compounds\cite{park2006hidden}. 
Due to its relevance to many experimental systems, 
intensive analytical
\cite{
doi:10.1080/0001873021000057123, 
PhysRevLett.84.5608, 
PhysRevLett.93.255702, 
PhysRevB.82.075128, 
PhysRevB.84.125115,
Abrahams28022012,
deCarvalho2013738,
PhysRevB.87.045104,
PhysRevB.91.125136, 
PhysRevB.93.165114, 
PhysRevB.92.165105,
PhysRevLett.115.186405}
and numerical 
\cite{
Berg21122012, 
2015arXiv151207257S, 
2015arXiv151204541L,
2016arXiv160908620G, 
2016arXiv160909568W,
Li2016925}
efforts have been made
to understand the nature of the non-Fermi liquid state.
The AF quantum critical metal in two dimensions
is described by a strongly interacting field theory
for the AF spin fluctuations and 
electronic excitations near the Fermi surface.
Although it seemed intractable,  
the theory for the $SU(2)$ symmetric 
AF quantum critical metal has been recently solved through
a non-perturbative ansatz\cite{2016arXiv160806927S}.
The non-perturbative solution utilizes a ratio between velocities, 
which dynamically flows to zero at low energies,
as a small parameter.

According to the non-perturbative solution\cite{2016arXiv160806927S},
the AF collective mode is strongly dressed by particle-hole excitations.
In contrast, electrons have zero anomalous dimension,
and exhibit a relatively weak departure from the Fermi liquid
with dynamical critical exponent $z=1$.
The non-perturbative solution actually applies to more general theories,
and the same conclusion holds for the AF quantum critical point 
in the presence of a one-dimensional Fermi surface 
embedded in general dimensions, $2 \leq d < 3$\cite{SDWgenD}.
However, the exact critical exponents obtained from the non-perturbative solution 
are not consistent with the earlier one-loop analysis 
of the theory in $3-\ep$ dimensions even in the small $\ep$ limit\cite{PhysRevB.91.125136}.
At the one-loop order, 
the ratio between velocities 
which is used as a small parameter 
in the non-perturbative solution does not flow to zero, 
and the electrons at the hot spots
exhibit a stronger form of non-Fermi liquid
with $z >1$.

In this work, we resolve this tension.
We extend the earlier one-loop analysis
to include higher-loop effects.
We find that the $\ep$-expansion
is not simply organized by the number of loops,
and certain higher-loop diagrams are enhanced by
IR singularities caused by an emergent quasi-locality.
As a result, a two-loop diagram qualitatively modifies the nature of the fixed point
even to the leading order in $\ep$\cite{PhysRevB.94.195135}.
We show that the $\ep$-expansion is controlled
with the inclusion of the two-loop effect.
Furthermore, the ratio between velocities
is shown to flow to zero in the low energy limit,
which protects the critical exponents from receiving
higher-loop corrections.
This is similar to the nematic critical point 
in $d$-wave superconductors, 
where an emergent anisotropy in velocities 
leads to asymptotically exact results
to all orders in the $1/N$ expansion\cite{PhysRevB.78.064512}.

The $\epsilon$-expansion and the non-perturbative solution\cite{2016arXiv160806927S,SDWgenD} are independent and complimentary.
The former is a brute-force perturbative analysis, which is straightforward but valid only near the upper critical dimension.
The latter approach is non-perturbative, and it is based on an Ansatz that is confirmed by a self-consistent computation.
The agreement of the results from the two different approaches provides an independent justification of the Ansatz used in the non-perturbative solution.

\section{Model and dimensional regularization}

\begin{figure}[h]
	\includegraphics[width=2.5in]{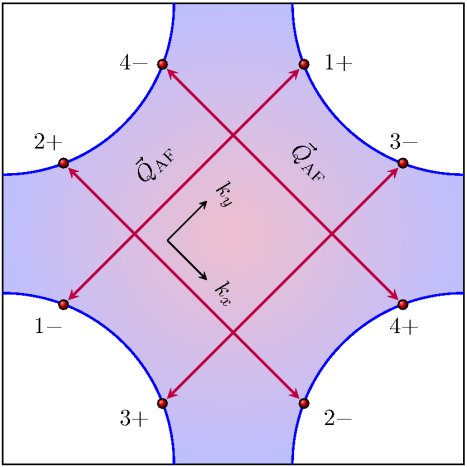}
	\caption{
		The first Brillouin zone of a metal in two dimensions  with $C_4$ symmetry.
		The shaded region represents the occupied states. 
		The AF ordering wavevector $\vec{Q}_{AF}$ is denoted by red arrows. 
		The hot spots are the red dots connected by $\vec{Q}_{AF}$.
	}
	\label{fig:hot spots}
\end{figure}

We start with the theory for the AF quantum critical metal in two dimensions. 
We consider a Fermi surface with the $C_4$ symmetry.
The low-energy degrees of freedom consist of 
the AF collective mode coupled to electrons near the hot spots, 
which are the set of points on the Fermi surface connected by the AF ordering vector \cite{PhysRevLett.84.5608, PhysRevLett.93.255702, PhysRevB.82.075128, PhysRevB.91.125136}, 
as is shown in Fig. \ref{fig:hot spots}. We study the minimal model, which has eight hot spots.
The AF ordering is taken to be collinear with a commensurate wave vector.
The action is written as
\onecolumngrid
\beqa
\mathcal{S} = &&  
\sum_{n=1}^4 \sum_{m=\pm} \sum_{\sigma=\uparrow,\downarrow} 
\int \f{d^3k}{(2\pi)^3} ~ 
{\psi}^{(m)*}_{n,\s}(k)
\left[ ik_0 ~ + e^{m}_n(\vec k;v)  \right] 
\psi^{(m)}_{n,\sigma}(k)
+ \half \int \frac{d^3q}{(2\pi)^3} 
\left[ q_0^2 + c^2 | \vec q |^2 \right] \vec{\phi}(-q)  \cdot \vec{\phi}(q)   \nn \\
&& + g_0 \sum_{n=1}^4 
\sum_{\s,\s'=\uparrow,\downarrow}
\int \f{d^3k}{(2\pi)^3} 
\f{d^3q}{(2\pi)^3} ~ 
\Bigl[
\vec{\phi}(q) \cdot 
{\psi}^{(+)*}_{n,\s}(k+q) \vec{\tau}_{\s,\s'}  
\psi^{(-)}_{n,\s'} (k) 
+ c.c. \Bigr] \nn \\
&& + \f{u_0}{4!} 
\int \f{d^3k}{(2\pi)^3}
\f{d^3p}{(2\pi)^3} 
\f{d^3q}{(2\pi)^3}
\lt[ \vec{\phi}(k+q) \cdot \vec{\phi}(p-q)\rt] 
\lt[ \vec{\phi}(-k) \cdot \vec{\phi}(-p) \rt]. 
\label{eq:3D_theory}
\eeqa
\twocolumngrid
\noindent
Here, $k = (k_0, \vec{k})$ denotes the Matsubara frequency and the two-dimensional momentum $\vec{k} = (k_x, k_y)$. 
$\psi_{n,\s}^{(m)}$ are the fermion fields that carry spin $\s = \uparrow, \downarrow$ at the hot spots labeled by $n=1,2,3,4, ~ m = \pm$. 
The choice of axis is such that the ordering wave vector is $\vec{Q}_{AF} = \pm \sqrt{2} \pi \hat{k}_x, \pm \sqrt{2} \pi \hat{k}_y$ up to the reciprocal lattice vectors $\sqrt{2} \pi (\hat{k}_x \pm \hat{k}_y)$. 
With this choice the fermion dispersions are 
$e^{\pm}_1(\vec k;v) = -e^{\pm}_3(\vec k;v) = v k_x \pm k_y$, 
$e^{\pm}_2(\vec k;v) = -e^{\pm}_4(\vec k;v) = \mp k_x + v k_y$, 
where $\vec{k}$ is the momentum deviation from each hot spot. 
The curvature of the Fermi surface can be ignored, 
since the patches of Fermi surface connected by 
the ordering vector are not parallel to each other 
with $v \neq 0$.
The Fermi velocity along the ordering vector has been set to unity by rescaling $\vec{k}$. 
$v$ is the component of Fermi velocity that is perpendicular to $\vec{Q}_{AF}$. 
$\vec{\phi}(q)$ is the boson field with three components
which describes the  AF collective mode 
with frequency $q_0$ and momentum $\vec{Q}_{AF} + \vec{q}$. 
$\vec{\tau}$ represents the three generators of the $SU(2)$ group. 
$c$ is the velocity of the AF collective mode. 
$g_0$ is the Yukawa coupling between the collective mode and the electrons near the hot spots,
and  $u_0$ is the quartic coupling between the collective modes.

We generalize the theory 
by tuning the number of co-dimensions 
of the one-dimensional Fermi surface\cite{PhysRevLett.102.046406, PhysRevB.88.245106,PhysRevB.91.125136}.
For this, we pair fermions on opposite sides of the Fermi surface into two component spinors,
 $\Psi_{1,\s} = (\psi_{1,\s}^{(+)}, \psi_{3,\s}^{(+)})^T$, 
 $\Psi_{2,\s} = (\psi_{2,\s}^{(+)}, \psi_{4,\s}^{(+)})^T$, 
 $\Psi_{3,\s} = (\psi_{1,\s}^{(-)}, - \psi_{3,\s}^{(-)})^T$, 
 $\Psi_{4,\s} = (\psi_{2,\s}^{(-)}, -\psi_{4,\s}^{(-)})^T$.
 In the spinor basis, the kinetic term for the fermions becomes 
 $S_F = \sum_{n=1}^4 \sum_{\sigma=\uparrow,\downarrow} \int \f{d^3k}{(2\pi)^3} ~ \bar{\Psi}_{n,\s}(k)
\left[ i \g_0 k_0 ~ + i \g_1 \eps_n(\vec{k};v)  \right] \Psi_{n,\sigma}(k)$, 
where $\g_0 = \s_y$ and $\g_1 = \s_x$ ($\s_i$ being the  Pauli matrices), $\bar{\Psi}_{n,\s} = \Psi^{\dagger}_{n,\s} \g_0$ 
with 
$\eps_1(\vec{k};v) = e_1^+(\vec{k};v)$, 
$\eps_2(\vec{k};v) = e_2^+(\vec{k};v)$, 
$\eps_3(\vec{k};v) = e_1^-(\vec{k};v)$, 
$\eps_4(\vec{k};v) = e_2^-(\vec{k};v)$. 
The general theory in $d$ spatial dimensions reads 
\onecolumngrid
\beqa
\mathcal{S} && \nn = 
\sum_{n=1}^4 
\sum_{\s=1}^{N_c} 
\sum_{j=1}^{N_f} 
\int dk~
\bar{\Psi}_{n,\s,j}(k) 
\Bigl[ i\mathbf{\Gamma} \cdot \mathbf{K}  
+ i \g_{d-1} \varepsilon_n(\vec{k};v) \Bigr] 
\Psi_{n,\s,j}(k)
+ \f14 
\int dq~
\Bigl[ \abs{\mathbf{Q}}^2 + c^2 | \vec q |^2 \Bigr]
\tr { \Phi(-q)  ~ \Phi(q) }
\\ \nn && + i \f{g \mu^{(3-d)/2}}{\sqrt{N_f}}
\sum_{n=1}^4
\sum_{\s,\s'=1}^{N_c}
\sum_{j=1}^{N_f} 
\int dk dq ~
\bar{\Psi}_{\bar n,\s,j}(k+q) 
\Phi_{\s,\s'}(q)
\g_{d-1} 
\Psi_{n,\s',j} (k) 
\\ && + \f{\mu^{3-d} }{4}
\int dk_1 dk_2 dq ~
\Bigl[ 
u_1 \tr { \Phi(k_1+q)  \Phi(k_2-q) } \tr { \Phi(-k_1)  \Phi(-k_2) }
+ u_2 \tr { \Phi(k_1+q)  \Phi(k_2-q)  \Phi(-k_1)  \Phi(-k_2) }
\Bigr]. 
\label{eq: generalD_theory}
\eeqa
\twocolumngrid
\noindent
Here we  consider $SU(N_c)$ spin and $N_f$ flavors of fermions for generality. 
$k = (\mathbf{K},\vec{k})$ is the $(d+1)$-dimensional energy-momentum vector
with $dk \equiv \f{d^{d+1} k}{(2\pi)^{d+1}}$.
$\vec{k} = (k_x, k_y)$ still denotes the two original momentum components, 
and  $\mathbf{K} = (k_0, k_1, ..., k_{d-2})$ denotes the frequency 
and the momentum components along the $(d-2)$ co-dimensions that have been added.
$\mathbf{\G} = (\g_0, \g_1, ..., \g_{d-2})$ together with $\g_{d-1}$ are the gamma matrices 
which satisfy the Clifford algebra $\{\g_{\mu},\g_{\nu}\} = 2 I \delta_{\mu,\nu}$ with $\Tr[I] = 2$. 
$\Psi_{n,\s,j}$ with $\s = 1, 2, ..., N_c$ and $j = 1, 2, ..., N_f$ is in the fundamental representation of
both the enlarged spin group $SU(N_c)$ and the flavor group $SU(N_f)$. 
$\Phi(q) = \sum_{a = 1}^{N_c^2 - 1} \phi^a(q) \tau^a$ is a matrix field for the collective mode, 
where $\tau^a$ are the generators of $SU(N_c)$ with $\Tr[\tau^a \tau^b] = 2 \delta_{ab}$. 
The Yukawa interaction scatters fermions between pairs of hot spots denoted  as $(n, \bar{n})$ with 
$\bar{1} = 3$, $\bar{2} = 4$, $\bar{3} = 1$, $\bar{4} = 2$. 
The Yukawa and quartic interactions have scaling dimensions $(3-d)/2$ and $(3-d)$, respectively,
at the non-interacting fixed point. 
$\mu$ is the energy scale introduced to make $g, u_1, u_2$ dimensionless.
For $N_c \leq 3$, $u_1$ and $u_2$ are not independent couplings
because of the identity, $\Tr[\Phi^4] = \f12 (\Tr[\Phi^2])^2$.
The energy of the fermions is given by 
$E_n(k_1, ..., k_{d-2}, \vec{k}) = \pm \sqrt{\sum_{i = 1}^{d-2} k_i^2 + \eps_n^2(\vec{k})}$,
which supports a one-dimensional Fermi surface
embedded in the $d$-dimensional momentum space. 
The theory respects 
the $U(1) \times SU(N_c) \times SU(N_f)$ internal symmetry.
It is also invariant under
the $C_4$ transformations in the $(k_x,k_y)$ plane, 
the $SO(d-1)$ that rotates $(k_0,...,k_{d-2})$,
and time-reversal. 
When $N_c = 2$, there is an additional pseudospin symmetry, 
which rotates $\Psi_{n,\s,j}(k)$ into $i \tau_{\s,\s'}^{(y)} \bar{\Psi}^T_{n,\s',j}(-k)$\cite{PhysRevB.82.075128}.

In three spatial dimensions the interactions are marginal. 
We therefore expand around $d = 3$ using $\ep = 3 - d$ as a small parameter. 
We use the minimal subtraction scheme to compute the beta functions, 
which dictate the renormalization group (RG) flow of the velocities and couplings. 
To make the quantum effective action finite in the ultraviolet (UV), 
we add counter terms which can be written in the following form,
\onecolumngrid
\beqa
	\mc{S}_{CT} \nn &=& 
	\sum_{n=1}^4 
	\sum_{\s=1}^{N_c} 
	\sum_{j=1}^{N_f} 
	\int dk ~
	\bar{\Psi}_{n,\sigma,j}(k) 
	\lt[ i \mc{A}_1 \mathbf{\Gamma} \cdot \mathbf{K}  
	+ i \mc{A}_{3} 
	\gamma_{d-1} 
	\eps_n\lt( \vec k; \fr{ \mc{A}_{2} }{ \mc{A}_{3} } v \rt)
	\rt] 
	\Psi_{n,\s,j}(k)
	\\ \nn &&  
	+ \f14 
	\int dq ~
	\Bigl[ \mc{A}_4 \abs{\mathbf{Q}}^2 + \mc{A}_5
	~ c^2 \abs{\vec q}^2 \Bigr]
	\tr{ \Phi(-q)  ~ \Phi(q) } 
	\\ \nn && 
	+ i \mc{A}_6 
	\f{g \mu^{(3-d)/2}}{\sqrt{N_f}}
	\sum_{n=1}^4
	\sum_{\s,\s'=1}^{N_c}
	\sum_{j=1}^{N_f} 
	\int dk \, dq ~  
	\Bigl[
	\bar{\Psi}_{\bar n,\s,j}(k+q) 
	\Phi_{\s,\s'}(q)
	\g_{d-1} 
	\Psi_{n,\s',j} (k) 
	\Bigr]  
	\\ \nn &&  +  \f{ \mu^{3-d} }{4}
	\int 
	dk_1 \, dk_2 \, dq ~
	\Bigl[ 
	\mc{A}_7 u_1 \tr { \Phi(k_1+q)  \Phi(k_2-q) } \tr{ \Phi(-k_1)  \Phi(-k_2) } 
	\\ && + \mc{A}_8 u_2 \tr { \Phi(k_1+q)  \Phi(k_2-q)  \Phi(-k_1)  \Phi(-k_2) } 
	\Bigl], 
	\label{eq: CT_action_general}
\eeqa
\twocolumngrid
where 
\eqaln{
	\mc{A}_n \equiv \mc{A}_n(v,c,g,u;\ep) = \sum_{m=1}^{\infty} \f{Z_{n,m}(v,c,g,u)}{\ep^m}.
	\label{eq:AZ}
}
$Z_{n,m}(v,c,g,u)$ are finite functions of the couplings. 
The counter terms are computed order by order in $\ep$.
The general expressions for the dynamical critical exponent, 
the anomalous scaling dimensions of the fields, 
and the beta functions of the velocities and couplings 
are summarized in Section III.
More details on the RG procedure can be found in Ref. \cite{PhysRevB.91.125136}.

\section{The modified one-loop fixed point}

\begin{figure}[h]
	\centering 
	\subfigure[]{\includegraphics[width=1.6in]{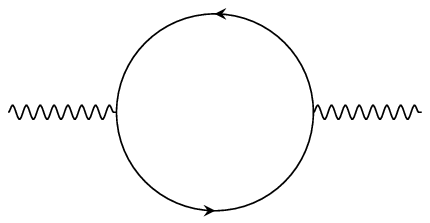}}
	\subfigure[]{\includegraphics[width=1.6in]{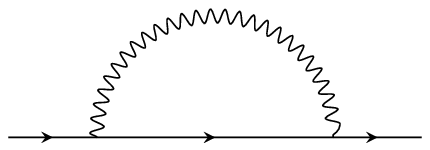}}
	\subfigure[]{\includegraphics[width=1.4in]{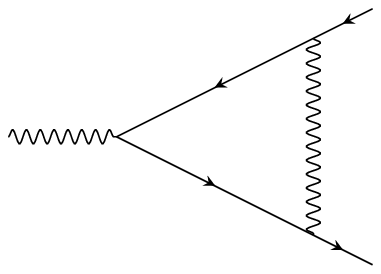}}
	\hspace{10mm}
	\subfigure[]{\includegraphics[width=1.0in]{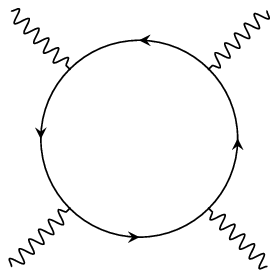}}
	\subfigure[]{\includegraphics[width=1.6in]{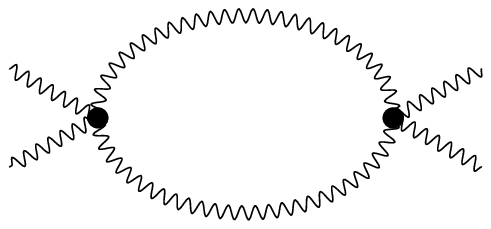}}
	\caption{One-loop diagrams.}
		\label{fig:one-loop diagrams}
\end{figure}

We begin by reviewing
the one-loop RG analysis
of Ref. \cite{PhysRevB.91.125136}.
The conclusion of the analysis is that
the theory flows to a quasi-local non-Fermi liquid state,
where $c, v$ flow to zero as $1/l$ for $d < 3$ and as $1/\log(l)$ at $d = 3$
in the logarithmic length scale $l$, 
with their ratio fixed to be $w \equiv v/c= \frac{N_c N_f}{N_c^2-1}$ in the low energy limit with $l \rightarrow \infty$.
Along with the emergent quasi-locality,
the couplings also flow to zero 
such that
$\lambda \equiv g^2/v$ and $\kappa_i \equiv u_i/c^2$ 
flow to  
$\l^* = \f{ 4 \pi (N_c^2+N_c N_f -1) }{ N_c^2+N_c N_f -3 }  \epsilon$
and $\kappa_i^* = 0$
in the low energy limit.

The perturbative expansion is controlled by the ratios 
between the couplings and the velocities,
and the dynamical critical exponent becomes
$z =  1 + \frac{N_c^2 + N_c N_f  - 1}{2(N_c^2 + N_c N_f - 3)} \epsilon$.
With $w \sim \mathcal{O}(1)$ at the one-loop fixed point,
general diagrams are estimated to scale as 
$I \sim \l^{\f{V_g}{2}} \k_i^{V_u} c^{V_u - L_b + \f{E-2}{2}}$,
where 
$V_g$ is the number of Yukawa vertices, 
$V_u$ is the number of quartic vertices, 
$L_b$ is the number of boson loops, and 
$E$ is the number of external lines.
Because $c$ flows to zero,
magnitudes of higher-loop quantum corrections are controlled not only by 
$\lambda$ but also by $c$.
In particular, the quantum correction to the spatial part of the boson kinetic term becomes
$\mc{A}_5 \sim I/c^2 \sim \l^{\f{V_g}{2}} ~ \k_i^{V_u} ~  c^{V_u - L_b - 2}$,
where the counter term is further enhanced by a factor of $1/c^2$ 
because the velocity in the classical action is already small.

\begin{figure}[h]
	\centering 
\includegraphics[width=2in]{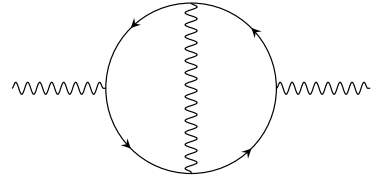}
	\caption{Two-loop diagram for the boson self-energy. }
	\label{fig:two-loop boson self energy}
\end{figure}

In three dimensions ($\ep=0$), 
all higher-loop diagrams are suppressed
because $\lambda$ flows to zero faster ($\lambda \sim 1/l$) 
than the velocities ($v \sim c \sim 1/\log(l)$).
Therefore, the critical point in three dimensions
is described by the stable quasi-local marginal Fermi liquid \cite{varma1989phenomenology},
where the Fermi liquid is broken by logarithmic corrections
from the one-loop effect\cite{PhysRevB.91.125136}.
Below three dimensions ($\ep >0$), however,
some higher-loop diagrams cannot be ignored
because $c$ flows to zero while $\lambda^* \sim \ep$. 
For example, $\mc{A}_5$ 
from Fig. \ref{fig:two-loop boson self energy} 
is divergent at the one-loop fixed point.
It might seem strange that 
the higher-loop graph suddenly becomes important for any nonzero $\ep$
while it is negligible at $\ep=0$.
This apparent discontinuity originates from the fact that
the small $\ep$ limit and the low energy limit do not commute.
If the small $\ep$ limit is taken first, all higher-loop graphs are suppressed.
However, since we are ultimately interested in the theory at $d=2$, 
we fix $\epsilon$ to a small but finite value, and then take the low energy limit of the corresponding theory.
In this case, $c$ flows to zero,
and the IR singularity caused by the softening of the collective mode 
enhances the magnitude of the two-loop graph.
Since certain higher-loop diagrams can be enhanced by the IR singularity in the small $c$ limit,
we cannot ignore all higher-order quantum corrections from the outset even in the small $\ep$ limit.

The largest contribution to the renormalization of $c$ comes from the boson self-energy 
in Fig. \ref{fig:two-loop boson self energy}. 
We call the addition of this two-loop diagram 
to the one-loop diagrams (Fig. \ref{fig:one-loop diagrams})
the ``modified-one-loop'' (M1L) order. 
As will be shown later, the flow of $c$ is modified by the two-loop graph in Fig. \ref{fig:two-loop boson self energy}
such that the effect of other higher-loop diagrams is negligible 
in the small $\ep$ limit.
There also exists a two-loop diagram made of quartic vertices contributing to $\mathcal{A}_5$.
However, the diagram has no enhancement by $1/c^2$ 
because the momentum dependent self-energy comes with $c^2$ 
due to the $(d+1)$-dimensional rotational symmetry present in the bosonic sector.
The contribution from the quartic vertices are further suppressed because
$\k_i$ is irrelevant at the fixed point.

Fig. \ref{fig:two-loop boson self energy} 
gives rise to the quantum effective action
whose divergent part is given by 
\beq
\delta \G_{0,2}^{2L} = 
		\f{1}{\ep} \f{4}{N_c N_f} \f{g^4}{v^2 c^2} h_5(v,c)
	\int dp ~
	\f14 c^2 |\vec{p}|^2 \text{Tr}[\Phi (-p)\Phi (p)],
	\label{eq:boson self-energy correction}
\eeq
where $h_5(v,c)$ is given by $h_5(v,c) = h_5^* \frac{v}{c}$
with $h_5^* \approx 5.7 \times 10^{-4}$
in the limit $v, c, v/c$ are small. 
The full definition of $h_5(v,c)$ is given in Appendix B. 
The positive sign of Eq. (\ref{eq:boson self-energy correction})
implies that the two-loop correction prevents $c$ 
from flowing to zero too fast\cite{PhysRevB.94.195135}.
If $c$ is small, the quantum correction makes the collective mode
speed up until the quantum correction becomes $\mathcal{O}(1)$, 
$\f{1}{\ep} \f{4}{N_c N_f} \f{g^4}{v^2 c^2 } h_5(v,c) \sim 1$.
Since $\f{g^2}{v} \sim \ep$, 
this suggests that $\f{g^2 }{c^3}$ becomes $\mathcal{O}(1)$ in the low energy limit.
Once $c$ becomes comparable to $g^{2/3} \sim v^{1/3}$,
it flows to zero together with $v$, 
although at a slower rate than $v$.
As a result, $w=v/c$ flows to zero at the M1L fixed point 
for $\ep>0$, unlike at $\ep=0$.
This emergent hierarchy in the velocities 
plays a crucial role in the non-perturbative solution\cite{2016arXiv160806927S, SDWgenD}. 
In order to confirm this picture,
we examine the RG flow in the space of $\{\l, x, w, \k_i\}$,
where $x \equiv  \f{g^2}{c^3}$ is expected to flow to an $\mathcal{O}(1)$ value at the fixed point.
%

The beta functions for  the five parameters are expressed in terms of the counter terms as
\beqa
\nn 
\f{d\l}{dl} &=&   z ~ \l \lt( \ep +  Z_{2,1}^\prime + Z_{3,1}^\prime + Z_{4,1}^\prime - 2 Z_{6,1}^\prime \rt),
\\ \nn
\f{d x}{dl} &=&   z ~ x \lt( \ep + \f12 
\lt( 6 Z_{1,1}^\prime - 2 Z_{3,1}^\prime - Z_{4,1}^\prime + 3 Z_{5,1}^\prime - 4 Z_{6,1}^\prime \rt)\rt),
\\ \nn
\f{dw}{dl} &=&   \f12~ z ~ w
\lt( 2Z_{1,1}^\prime -2 Z_{2,1}^\prime - Z_{4,1}^\prime + Z_{5,1}^\prime \rt),
\\ \nn
\f{d\k_1}{dl} &=&   z ~ \k_1 \lt( \ep + Z_{4,1}^\prime + Z_{5,1}^\prime - Z_{7,1}^\prime \rt),
\\
\f{d\k_2}{dl} &=&   z ~ \k_2 \lt( \ep + Z_{4,1}^\prime + Z_{5,1}^\prime - Z_{8,1}^\prime \rt),
\label{eq:beta functions for w lambda x kappa}
\eeqa
where 
$Z_{n,1}^{\prime} \equiv \lt(\f12 g \partial_g 
+ u_i \partial_{u_i} \rt)  Z_{n,1}$,
and
$z = \lt[ 1 + Z_{1,1}^\prime - Z_{3,1}^\prime  \rt]^{-1}$
is the dynamical critical exponent.
In the limit that $v,c,v/c$ are small, 
the beta functions at the M1L level become
\onecolumngrid
\beqa 
	\f{d\l}{dl} &=&   z ~ \l \Bigg(\ep - \f{1}{4 \pi} \l + \f{1}{2\pi N_c N_f}  ~ \l w
	\Bigg),
	\label{eq:lambda beta function}
	\\ 
	\f{dx}{dl} &=&  z ~ x ~
	\Bigg(
	\ep   
	- \f{3 N_c^2 - 7}{8 \pi N_c N_f } ~ \l w 
	+ \f{(N_c^2 - 1)}{2 \pi^2 N_c N_f }  ~ \f{(\l w)^{\f32}}{x^{\f12}} 
	+ \frac{1}{8 \pi} \l
	- \f{12 \, h_5^*}{N_c N_f} ~ \l \, x
	\Bigg),
	\label{eq:x beta function}
	\\
	\f{dw}{dl} &=&   \f12~ z ~ w 
	\Bigg(
	- \f{(N_c^2 - 1)}{4 \pi N_c N_f }  ~ \l w  
	- \f{(N_c^2 - 1)}{\pi^2 N_c N_f } 
	~ \f{(\l w)^{\f32}}{x^{\f12}} 
	+ \f{1}{4 \pi} \l
	- \f{8 \, h_5^*}{N_c N_f} ~ \l \, x
	\Bigg),
	\label{eq:w beta function}
	\\ 
	\f{d\k_1}{dl} &=&   z ~ \k_1 \lt( \ep 
	-\frac{1}{4 \pi} \l - \f{8 \, h_5^*}{N_c N_f} ~ \l \, x - \frac{1}{2 \pi^2} 
	\lt( (N_c^2+7)\k_1 
	+ 2 \lt( 2 N_c - \frac{3}{N_c} \rt) \k_2 
	+ 3 \lt( 1 + \frac{3}{N_c^2} \rt) \frac{\k_2^2}{\k_1} \rt)
	\rt),
	\label{eq:kappa_1 beta function}
	\\ 
	\f{d\k_2}{dl} &=&   z ~ \k_2 \lt( \ep 
	-\frac{1}{4 \pi} \l - \f{8 \, h_5^*}{N_c N_f} ~ \l \, x - 
	\frac{1}{2 \pi^2} \lt( 12 \k_1 + 2 \lt( N_c - \frac{9}{N_c} \rt) \k_2  \rt)
	\rt),
	\label{eq:kappa_2 beta function}
\eeqa
\twocolumngrid
\noindent
with $z = \left( 1 - \frac{N_c^2 - 1}{8 \pi N_c N_f} \lambda w \right)^{-1}$.
The beta functions exhibit a stable fixed point given by
\beq
	\l^* = 4 \pi \ep, \quad x^* = \f{N_c N_f}{32 \pi \, h_5^*}, \quad w^* = 0, \quad \k_i^* = 0.
	\label{eq:fixed point values}
\eeq
It is noted that $x$ is $\mathcal{O}(1)$, and $v,c,v/c$ all vanish at the fixed point.

\begin{figure}[h]
	\centering 
	\includegraphics[width=2.5in]{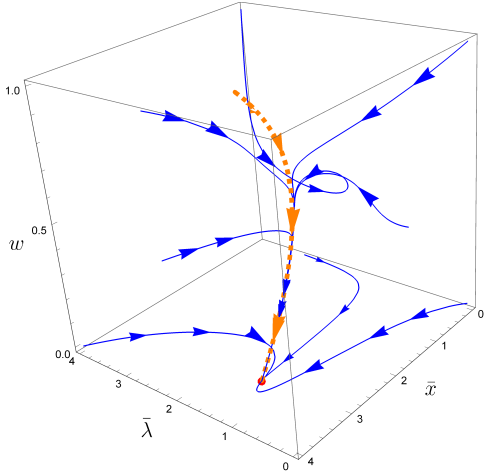}
	\caption{
The RG flow in the space of 
$(\lambda, x, w)$ 
for $\ep = 0.01$ and $N_c = 2, N_f = 1$. 
The axes are scaled as $\bar{x} \equiv x/10$, $\bar{\lambda} \equiv 10 \lambda$.
The fixed point 
$(\bar{\lambda}^*, \bar{x}^*, w^*) = ( 1.26, 3.49, 0)$ 
is denoted as a red dot. 
The solid curves represent the numerically integrated RG flows,
and the dotted (orange) line represents the one-dimensional manifold 
given by $\frac{d x}{dl }= \frac{d \l}{dl}=0$.
	}
	\label{fig:numerical RG flow}
\end{figure}

In order to understand the flow near the fixed point,
we first examine the beta functions for  $x$ and $\lambda$.
Although it may seem arbitrary to focus on the flow of $x,\l$ first with fixed $w$,
this is actually a good description of the full RG flow because
the flow of $x,\l$ is much faster than that of $w$, as will be shown in the following.
From Eqs. (\ref{eq:lambda beta function}), (\ref{eq:x beta function}),
the beta functions for $(\delta \l, \delta x) \equiv (\l - \l^*, x - x^*)$
are given by
\beqa
\f{d \delta \l}{dl} &=&    
 f_{\l}(w)
- \ep ~ \delta \l + \dots,
\nn
\\ 
\f{d \delta x}{dl} &=&   
 f_{x}(w)
- \f{N_c N_f}{32 \pi \, h_5^*} \,
\Bigg(
     \f{48 \pi \, h_5^* \ep \delta x}{N_c N_f} 
+  \frac{\delta \l }{4 \pi} \Bigg)
+ \dots
 \nn \\
\label{eq:x beta function lowest order}
\eeqa
to the linear order in the deviation from the fixed point for small $w$, 
where 
$f_{\l}(w) = \left. \frac{d \l}{d l} \right|_{\l = \l^*, x=x^*}$,
$f_{x}(w) = \left. \frac{d x}{d l} \right|_{\l = \l^*, x=x^*}$,
and 
$\dots$ represent terms that are higher order in $\delta \lambda$, $\delta x$.
Eq. (\ref{eq:x beta function lowest order}) implies that
the perturbations in $\lambda$ and $x$
are irrelevant at the fixed point, 
and they flow to $w$-dependent values exponentially in $l$. 
This can be seen from Fig. \ref{fig:numerical RG flow},
which shows the full numerical solution 
to the beta functions for $(\lambda, x, w)$.
Once the RG flow reaches the one-dimensional manifold  given by
$
(\lambda, x, w)
= \left( 
\l^* + \frac{ f_{\l}(w) }{\ep},
x^* + \frac{2}{3 \ep} 
\left[ f_x(w) - \f{N_c N_f f_{\l}(w) }{128 \pi^2 \, h_5^* \ep} \right], 
w 
\right)$,
$w$ flows to the fixed point at a slower rate.
To compute the flow within this manifold,
we set $\f{d  \l}{dl} = \f{d x}{dl} = 0$
in Eq. (\ref{eq:beta functions for w lambda x kappa})
to express  
$ Z_{1,1} $,
$ Z_{4,1} $
in terms of
$ Z_{n,1} $
with $n=2,3,5,6$.
This gives the beta function for $w$ within the manifold,
\beqa
\f{dw}{dl} &=&   \frac{2 z }{3} 
\lt( -Z_{2,1}^{'} + Z_{3,1}^{'} \rt) w,
\eeqa
which reduces to
\beq
\f{d w}{dl} = -   \f{64 \sqrt{2 h_5^*} (N_c^2 - 1)}{3 (N_c N_f)^{3/2}} \ep^{3/2} \, w^{5/2}
\label{eq:beta function w low energy}
\eeq
to the leading order in $w$.
Because the flow velocity of $ w$ vanishes to the linear order in $w$, 
$w$ flows to zero as a power-law in the logarithmic length scale, $w \sim l^{-2/3}$. 
At the fixed point, the quartic couplings are irrelevant
and their beta functions become 
\beqa
\f{d \k_i}{dl} &=& - \ep  \k_i, 
\label{eq:beta functions kappa linear order}
\eeqa
to the leading order in $w$ and $\k_i$.
This confirms that the fixed point in Eq. (\ref{eq:fixed point values}) is stable.

In the small $\ep$ limit,
Eq. (\ref{eq:fixed point values}) does not converge 
to the one-loop fixed point,
$\lambda^* = 0$, $x^* = 0$, 
$w^* = \f{N_c N_f}{N_c^2-1}$, 
$\k_i^* = 0$, 
which represents the correct fixed point at $\ep=0$.
Although the beta functions are analytic functions of $\ep$, 
the fixed point is not
because the low energy limit and the $\ep \rightarrow 0$ limit do not commute.
One way to understand this non-commutativity 
is in terms of  the `RG time' that is needed
for the flow to approach Eq. (\ref{eq:fixed point values})
for nonzero but small $\ep$.
In order for $w$ to decrease by a factor of $1/2$,
the logarithmic length scale has to change by
$\Delta l \sim \ep^{-3/2}$
according to Eq. (\ref{eq:beta function w low energy}).
The fixed point described by Eq. (\ref{eq:fixed point values})
can be reached only below the crossover energy scale,
$\mu \sim \Lambda e^{- \ep^{-3/2}}$,
where $\Lambda$ is a UV cut-off scale.
The crossover energy scale goes to zero as $\ep$ becomes smaller,
and the fixed point in Eq. (\ref{eq:fixed point values})
is never reached at $\ep=0$.
A converse issue of non-commutativity arises in  
$2 + \ep^\prime$ dimensions\cite{SDWgenD}.
In order to capture the correct physics in two dimensions, one needs to take the $\ep^\prime \rightarrow 0$ limit
first before taking the low energy limit.
If the other order of limits is taken,
some logarithmic corrections are missed\cite{SDWgenD}.

Although the two-loop diagram in Fig. \ref{fig:two-loop boson self energy} (a) 
is superficially $\mathcal{O}(\ep^2)$,
it becomes $\mathcal{O}(\ep)$ at the fixed point because
the IR singularity caused by the vanishingly small velocities
enhances the magnitude of the diagram.
Formally, a factor of $g^2$ coming from one additional loop
is canceled by an IR enhancement of $c^{-3}$ in Eq. (\ref{eq:boson self-energy correction}),
which makes the two-loop diagram 
as important as the one-loop diagrams in the small $\ep$ limit.
This is rather common in field theories of Fermi surfaces
where the perturbative expansion is not 
organized by the number of loops\cite{PhysRevB.80.165102,PhysRevB.82.075128,PhysRevB.88.245106,PhysRevB.94.195135}.

The breakdown of the naive loop expansion is analogous to 
the case of the ferromagnetic quantum critical point\cite{RevModPhys.88.025006}.
In the disordered  ferromagnetic quantum critical metal, the perturbative expansion breaks down even near the upper critical dimension, as a dangerously irrelevant operator enters in the beta functions of other couplings in a singular manner\cite{PhysRevB.63.174428,PhysRevLett.93.155701}. 
In our case, the velocities play the role of dangerously irrelevant couplings which spoil the naive loop expansion.
Although they are marginally irrelevant, one cannot readily set the velocities to zero as quantum corrections are singular in the zero velocity limit. 
This leads to a subtle balance between the Yukawa coupling and the velocities, 
making the two-loop diagram as important as the one-loop diagrams.
Then the natural question is the role of  other higher-loop diagrams. 
In the following,  we show that other higher-loop diagrams are suppressed
and the $\epsilon$-expansion is controlled, 
as is the case for the SDW critical metal with $C_2$ symmetry\cite{PhysRevB.94.195135}.

\section{Emergent small parameter}

In this section, 
we show that the $\ep$-expansion is controlled, 
by providing an upper bound for the magnitudes 
of general higher-loop diagrams at the M1L fixed point.
Furthermore, we show that a large class of diagrams are further suppressed by
$w$, which flows to zero in the low energy limit.
Since $\kappa_i=0$ at the M1L fixed point, 
only those diagrams without quartic vertices are considered.
Among the diagrams made of only Yukawa vertices,
we first focus on the diagrams without self-energy corrections. 
The diagrams without self-energy corrections scale as
\beqa
I \sim \f{g^{2L+E-2}}{v^{L_f}c^{L - L_f}},
\label{eq:scaling v c Yukawa only}
\eeqa
up to potential logarithmic corrections in $v$ and $c$,
where  $L$ is the total number of loops,
$L_f$ is the number of fermion loops,
and $E$ is the number of external lines.
The derivation of Eq. (\ref{eq:scaling v c Yukawa only}),
which closely follows Ref. \cite{2016arXiv160806927S},
can be found in Appendix C.

A diagram whose overall magnitude is given by Eq. (\ref{eq:scaling v c Yukawa only}) 
contributes to the counter term as
\beqa
\mathcal{A}_1, \mathcal{A}_2, \mathcal{A}_3, \mathcal{A}_4, \mathcal{A}_6  & \sim &  
\lambda^{L} ~w^{L-L_f},  \nn \\
\mathcal{A}_5 & \sim &  \lambda^{L-1} ~w^{L-L_f-1} ~x,
\label{eq:As}
\eeqa
up to logarithmic corrections in $v$ and $c$,
where the relations,
$g = \left( \frac{ \lambda^3 w^3}{x} \right)^{\f14}$,
$v = \left( \frac{ \lambda w^3}{x} \right)^{\f12}$ and
$c = \left( \frac{ \lambda w}{x} \right)^{\f12}$ are used. 
$ \mathcal{A}_5$ scales differently from the rest of the counter terms  
because quantum corrections to the spatial part of the boson kinetic term
are enhanced by $\frac{1}{c^2}$.
Since the classical action $c^2 |\vec q|^2$ vanishes in the $c \rightarrow 0$ limit,
the relative magnitude of quantum corrections to the classical action is enhanced 
as $ \mathcal{A}_5 \sim  \frac{1}{c^2} I$.
For example, the two-loop diagram in Fig. \ref{fig:two-loop boson self energy}
is $ \mathcal{A}_5 \sim  \frac{g^4}{v c^3}$.
On the other hand, $ \mathcal{A}_2$ is not enhanced by $\f1v$, 
even though the fermion kinetic term also loses its dependence
on $k_x$ ($k_y$) for $n=1,3$ ($n=2,4$)
in the small $v$ limit.
The difference is attributed to the fact that the fermion self-energy takes the form of 
$ \Sigma(k) \sim  \f{g^{2L}}{v^{L_f}c^{L - L_f}} \tilde \Sigma( k_0, v k_x, k_y)$  for $n=1,3$
and 
$ \Sigma(k) \sim  \f{g^{2L}}{v^{L_f}c^{L - L_f}} \tilde \Sigma( k_0, k_x, v k_y)$  for $n=2,4$.
Besides the overall factor of  $\f{g^{2L}}{v^{L_f}c^{L - L_f}}$ 
from Eq. (\ref{eq:scaling v c Yukawa only}),
$\tilde \Sigma$ becomes independent of 
$k_x$ ($k_y$) for $n=1,3$ ($n=2,4$) in the small $v$ limit. 
This is because in all fermion self-energy diagrams
the external momentum can be directed to flow through 
a series of fermion propagators of type $n=1,3$ ($n=2,4$) only,
and the fermion propagators become independent of $k_x$ ($k_y$) when $v=0$.
For example, the one-loop fermion self-energy 
with $L=1, L_f=0$ in Fig. \ref{fig:one-loop diagrams}
is at most $\Sigma \sim \frac{g^2}{c} ( v k_x - k_y )$ for $n=1$.
Explicit calculation actually shows that the one-loop diagram
is further suppressed by $c$ 
for an unrelated reason\cite{PhysRevB.91.125136}.

\begin{figure}[h]
	\centering 
	\includegraphics[width=3.5in]{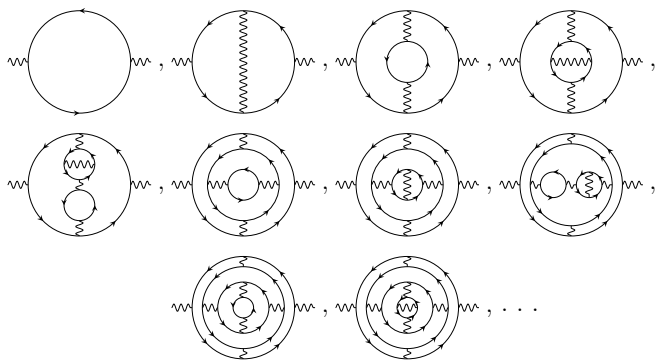}
	\caption{
Some examples in the infinite series of diagrams 
that survive in the small $w$ limit.
	}
	\label{fig:boson self energy delta S^*}
\end{figure}
\twocolumngrid

Now we consider the consequences of Eq. (\ref{eq:As}). 
We initially ignore the potential logarithmic 
corrections in $v,c$.
First, higher-loop diagrams are systematically suppressed by $\lambda^* \sim \ep$
as the number of loops increases.
However, there is an exception to the usual rule 
that $L$-loop diagrams are suppressed  by $\ep^{L}$. 
The quantum correction to the spatial part of the boson kinetic term
is suppressed only by $\ep^{L-1}$,
due to the enhancement by $1/c^2$.
Although Eq. (\ref{eq:As}) suggests that the one-loop contribution to $\mathcal{A}_5$ 
scales as $\lambda^0 w^{-1} x$,
its contribution to $\mathcal{A}_5$ is actually zero
because Fig. \ref{fig:one-loop diagrams} (a)  
is independent of momentum.
Since all self-energy corrections are at most $O(\ep)$, 
diagrams with  self-energy insertions are further suppressed by $\ep$.
This implies that the $\ep$-expansion is controlled, 
and the M1L includes all quantum corrections to the linear order in $\ep$.

Second, a large class of higher-loop diagrams are further suppressed by $w$
which flows to zero in the low energy limit.
Unlike $\ep$, which is fixed at a given dimension,
$w$ flows to zero dynamically in the low energy limit.
The suppression by $w$ is controlled by the number of non-fermion loops.
The only diagram with $L-L_f=0$ is the one-loop boson self-energy in Fig. \ref{fig:one-loop diagrams} (a).
Since $\mathcal{A}_5$ 
from Fig. \ref{fig:one-loop diagrams} (a) vanishes,
the leading order contribution  to $\mathcal{A}_5$ comes from
the two-loop boson self-energy in Fig. \ref{fig:two-loop boson self energy}
at $\mathcal{O}(w^0)$.
Among the diagrams without self-energy insertions,
only Fig. \ref{fig:one-loop diagrams} (a) and Fig. \ref{fig:two-loop boson self energy}
survive in the small $w$ limit.
When those self-energy corrections are included inside a diagram,
the diagram with dressed boson propagators
is not further suppressed by $w$ 
(although they are suppressed by $\ep$).
Other self-energy corrections, including all fermion self-energies, are negligible because they are suppressed by $w$.
Therefore, the complete set of diagrams which survive in the small $w$ limit
are generated by dressing the boson propagator
in Fig. \ref{fig:two-loop boson self energy}
by the self-energy in Fig. \ref{fig:one-loop diagrams} (a) and Fig. \ref{fig:two-loop boson self energy}.
This generates a series of diagrams, 
some of which are shown in Fig. \ref{fig:boson self energy delta S^*}.

Now we turn our attention to the sub-leading corrections 
that are potentially logarithmically divergent in $v$ and $c$ in Eq. (\ref{eq:As}). 
Diagrams suppressed by at least one power of $w$ still vanish in the small $w$ limit
even in the presence of logarithmic divergences in $v$ or $c$.
However, the effect of the logarithms on the diagrams in Fig. \ref{fig:boson self energy delta S^*} 
(which are $\mathcal{O}(w^0)$) cannot be ignored,
and this can in principle jeopardize the control of the $\ep$-expansion.
In Appendix D, we demonstrate that the $\ep$-expansion is still controlled,
by showing that all logarithmic corrections that arise at higher orders in $\ep$ can be absorbed into 
$\tilde x = x/F(c,v)$, 
where $F(c,v)$ is defined 
such that $\tilde x$ flows to $x^*$ 
in the low energy limit.
Once physical observables are expressed in terms of the new parameter $\tilde x$,
they have a well defined expansion in $\ep$.
At least for small $\ep$, 
the theory is free of perturbative instabilities
toward other competing orders
\cite{
PhysRevB.72.174520,
PhysRevB.82.075128,
Berg21122012,
Efetov2013,
PhysRevLett.117.157001,
PhysRevB.89.195115},
and it represents a stable 
non-Fermi liquid state\cite{PhysRevB.88.245106,PhysRevB.90.045121,PhysRevB.94.195135}.

\begin{figure}[h]
	\centering 
	\subfigure[]{\includegraphics[width=1.0in]{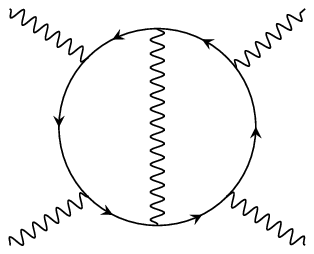}} \hspace{7mm}
	\subfigure[]{\includegraphics[width=1.0in]{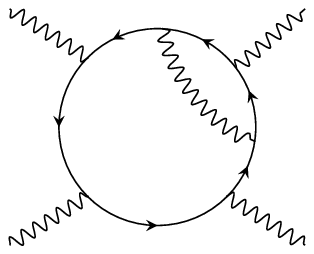}} \hspace{7mm}
\subfigure[]{\includegraphics[width=1.2in]{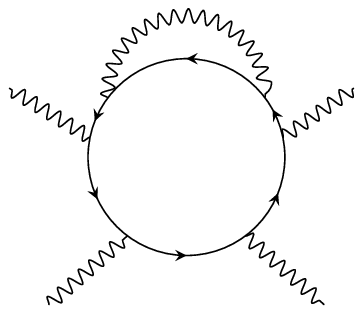}}
	\caption{
Quantum corrections that renormalize the quartic vertices in the small $w, \ep$ limit.
}
	\label{fig:two loop boson vertex}
\end{figure}

Although $\k_i=0$ at the M1L fixed point,
the quartic vertices are generated from the Yukawa vertices.
It happens that the one-loop diagram in Fig. \ref{fig:one-loop diagrams}(d) vanishes,
and the leading contributions that source the quartic vertices
are shown in Fig. \ref{fig:two loop boson vertex}. 
Once these diagrams are included, the beta functions for $\kappa_i$
are modified as
 \beqa
\f{d \k_i}{dl} &=& - \ep  \,  \k_i + A_i \lambda^{\f52} w^{\f32} x^{\f12}, 
\label{eq:beta functions kappa higher}
\eeqa
where the $A_i$'s are functions that diverge at most logarithmically in $w$ in the small $w$ limit.
As a result, the quartic couplings flow to zero as $\kappa_i \sim w^{\f32}$ 
up to logarithms of $w$ as $w$ flows to zero.

The small parameter $w$ that emerges in the low energy limit
suppresses all higher-loop diagrams except for the specific set of
diagrams shown in Fig. \ref{fig:boson self energy delta S^*}.
It turns out that $w$ flows to zero in the low energy limit in any dimensions, 
$2 \leq d < 3$\cite{2016arXiv160806927S,SDWgenD}.
This allows one to extract the exact critical exponents
by non-perturbatively summing the infinite series of diagrams through a self-consistent equation.


	\section{Physical Properties}

Now, we examine the scaling form of the Green's functions.
The dynamical critical exponent and the anomalous scaling dimensions at the fixed point are given by
	\beq
	z = 1, \quad
	\eta_{\psi} = 0, \quad
	\eta_{\phi} = \frac{\ep}{2}.
	\label{eq:critical exponents at fixed point}
	\eeq
These critical exponents do not receive higher-order corrections in $\ep$ in the small $w$ limit,
as is shown in Appendix D.
Indeed, $w$ flows to zero in the low energy limit,
and the critical exponents in Eq. (\ref{eq:critical exponents at fixed point})
are exact in any $0<  \ep \leq 1$\cite{2016arXiv160806927S,SDWgenD}.
At intermediate energy scales, 
the physical Green's functions receive 
corrections generated from irrelevant parameters of the theory.
The least irrelevant parameter that decays at the slowest rate is $w$,
which decays as $l^{-2/3}$  in the logarithmic scale $l$.
This sub-logarithmic flow introduces super-logarithmic corrections in the Green's functions. 
The fermion Green's function for the $n = 1$ patch is given by
\beqa 
 &&	G_1(\mathbf{K},\vec{k}) = \frac{1}{iF_{\psi}(|\mathbf{K}|)}
  \times \nn \\  && 
\frac{1}{	F_{z}(|\mathbf{K}|)	\boldsymbol{\Gamma}\cdot\mathbf{K}
     +\gamma_{d-1} \left[   \frac{\pi N_c N_f}{4\epsilon(N^2_c-1)}
	\frac{k_x}{\log(1/|\mathbf{K}|)}+k_y \right]   }    
	\label{eq:G_1 scaling}
\eeqa
	in the limit of small frequency $|\mathbf{K}|$ and fixed $e^{\mathfrak{I}_{z}(l)}\vec{k}\sim 1$, 
where $\mathfrak{I}_z(l) = l-\frac{3(N^2_c-1)^\frac{1}{3}}{2^\frac{14}{3}(h^{*}_5)^\frac{1}{3}}l^\frac{1}{3}$ and $l=\log(1/|\mathbf{K}|)$. 
The universal functions $F_z(|\mathbf{K}|)$ and $F_{\psi}(|\mathbf{K}|)$,
	\begin{align}
	F_z(|\mathbf{K}|)&=\exp\left(\frac{3(N^2_c-1)^\frac{1}{3}}{2^\frac{14}{3}(h^{*}_5)^\frac{1}{3}}\left(\log\frac{1}{|\mathbf{K}|}\right)^\frac{1}{3}\right),
	\label{eq:Fz}\\
	F_{\psi}(|\mathbf{K}|) &=\sqrt{\log\frac{1}{|\mathbf{K}|}},
	\label{eq:Fpsi}
	\end{align}
contain the contributions from the deviations of the dynamical critical exponent 
and the anomalous scaling dimension of the fermion, respectively, from their fixed point values in Eq. (\ref{eq:critical exponents at fixed point}).
Due to the super-logarithmic correction, the quasiparticle peak is destroyed.
All other Green's functions are determined by this one through the $C_4$ symmetry of the theory.

The scaling form of the spin-spin correlation function is given by
	\begin{align}
	& D(\mathbf{Q}, \vec{q}) 
	= \frac{1}{|\mathbf{Q}|^{2-\epsilon}F_{z}(|\mathbf{Q}|)^2F_\phi(|\mathbf{Q}|)} \nn \\
&	\times \mathfrak{D}\left(\frac{\vec{q}}{F_z(|\mathbf{Q}|)|\mathbf{Q}|}; \frac{N_c N_f}{2^\frac{11}{3}(h^{*}_5)^\frac{1}{3}(N^2_c-1)^\frac{2}{3}}\frac{1}{\epsilon }\frac{1}{\log(1/|\mathbf{Q}|)^\frac{2}{3}} \right),
	\label{eq:D scaling}
	\end{align}
	in the limit of small frequency $|\mathbf{Q}|$ and fixed $e^{\mathfrak{I}_{z}( \log(1/|\mathbf{Q}|) )}\vec{q}\sim 1$.
$\mathfrak{D}$ is a universal function, and 
	\begin{align}
	F_{\phi}(|\mathbf{Q}|)&=\exp\left( -\frac{3(N^2_c-3)}{2^\frac{11}{3}(h^{*}_5)^\frac{1}{3} (N^2_c-1)^\frac{2}{3}}\left(\log\frac{1}{|\mathbf{K}|}\right)^\frac{1}{3}\right)
	\label{eq:F_phi}
	\end{align}
	is the universal function 
which captures the contribution from the deviation of the anomalous scaling dimension of the boson field from its fixed point value in Eq. (\ref{eq:critical exponents at fixed point}). 
Unlike the fermion Green's function, the boson has a non-trivial anomalous dimension.

\section{Physical picture}

\begin{figure}[h]
	\centering 
	\subfigure[]{\includegraphics[width=3.in]{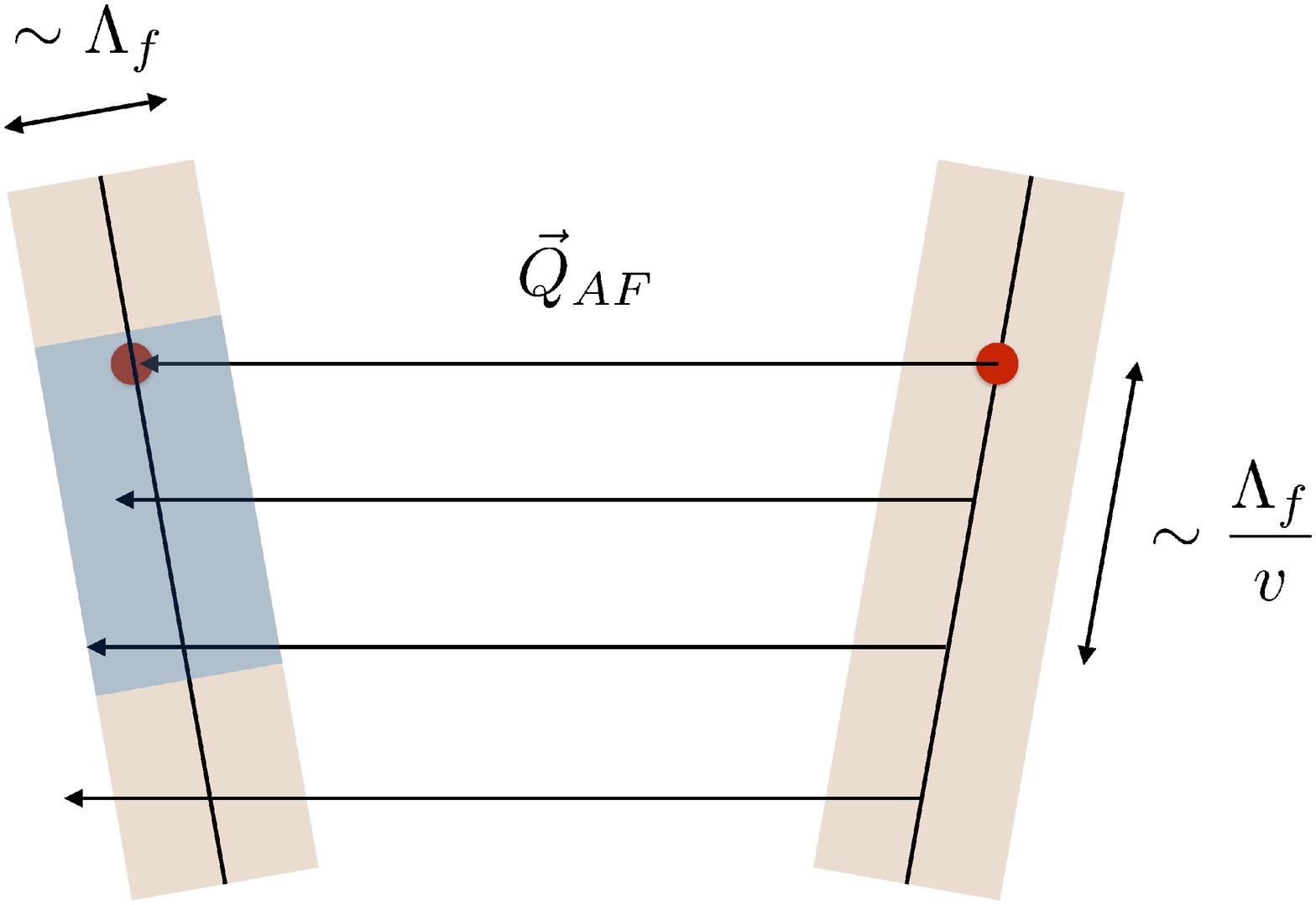}} 
	\subfigure[]{\includegraphics[width=3.in]{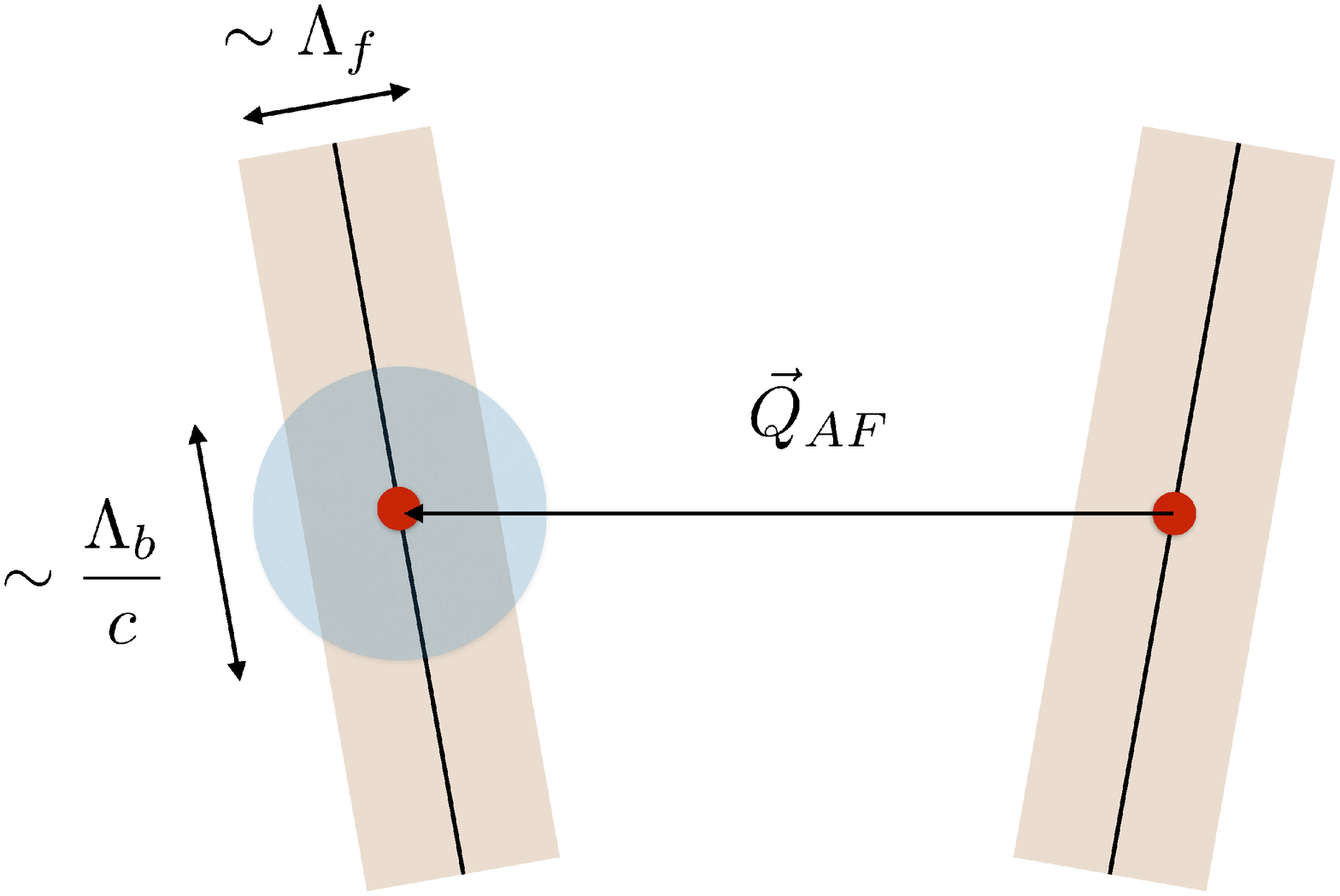}} 
	\caption{
The tilted lines represent patches of Fermi surface connected by the AF ordering vector, 
where the red dots denote hot spots.
The Fermi surfaces are not parallel because of non-zero $v$.
(a) Particle-hole excitations of momentum $\vec{Q}_{AF}$ 
can stay within the low-energy states of energy $E < \Lambda_f$
as far as their momenta are within the range of $\Lambda_f/ v$ from the hot spots.
Therefore the phase space available for Landau damping of the collective mode scales as $1/v$ in the small $v$ limit.
(b) The shaded region denotes the phase space available for a fermion 
when scattered by a collective mode of energy less than  $\Lambda_b$.
Since the energy of the boson with momentum $\vec q$ scales as $c |\vec q|$,
a boson with energy less than $\Lambda_b$ can transfer momentum up to $\Lambda_b/c$ 
to a fermion.
Therefore, the phase space grows as $1/c$ in the small $c$ limit.
}
	\label{fig:phase space for scattering}
\end{figure}

Finally, we provide a simple physical picture 
for why $w = v/c$ emerges as a control parameter.
The most important factor is the Landau damping
which describes the decay of the collective mode into the particle-hole continuum.
As the Fermi surface becomes locally nested near the hot spots in the small $v$ limit, 
the phase space for the particle-hole excitations that a collective mode can decay into increases as $1/v$. 
A single boson with a fixed momentum can decay into low-energy particle-hole pairs 
that lie anywhere along the nested Fermi surface of length $\Lambda_f/v$, 
where $\Lambda_f$ is an energy cut-off for the fermionic excitations. 
This is illustrated in Fig. \ref{fig:phase space for scattering}(a). 
This results in a large screening,
which renormalizes the Yukawa vertex to $g^2 \sim \ep v$.
As the Fermi surface gets nested, $g$ flows to zero.

The dispersionless particle-hole excitations near the hot spots 
renormalize the velocity of the collective mode to zero as well,
through the mixing between the collective mode and the particle-hole excitations.
As the fluctuations of the collective mode become soft,
quantum fluctuations are enhanced at low energies.
On the other hand, the enhanced quantum fluctuations speeds up the velocity of the collective mode
through Fig. \ref{fig:two-loop boson self energy}, 
and a balance is formed such that $c^3 \sim g^2 \sim \ep v$ to the leading order in $\ep$.
As a result, the boson velocity $c$ flows to zero at a much slower rate than $v$.

Now let us consider the feedback of the collective mode on the propagation of fermions,
by examining the process where a fermion is scattered by a collecitve mode. 
With the initial momentum fixed, the fermion does not have access to the entire Fermi surface.
Instead it can only scatter into a region allowed by the maximum momentum carried by a collective mode.
The available phase space for the scattering scales as
$\Lambda_b / c$,
where $\Lambda_b$ is the energy cut-off of the collective mode.
This is illustrated in Fig. \ref{fig:phase space for scattering}(b). 
Therefore, the scattering of fermions is controlled by $g^2/c \sim \ep \, v/c$, 
where $g^2 \sim \ep v$ is used.
As $v/c$ flows to zero in the low energy limit,
the scattering of fermions by collective modes becomes negligible.
This explains why fermions are largely intact
in the small $w$ limit,
and $w$ emerges as a control parameter.

\section{Conclusion}

We extended the earlier one-loop analysis of the
antiferromagnetic quantum critical metal
based on the dimensional regularization scheme
which tunes the number of co-dimensions 
of the one-dimensional Fermi surface.
We show that the IR singularities caused by the 
emergent quasi-locality rearrange the perturbative series
such that a two-loop graph becomes as
important as the one-loop graphs in the small $\ep$ limit.
With the inclusion of this two-loop effect,
higher-loop diagrams are systematically suppressed,
and the $\ep$-expansion is controlled.
Furthermore, a ratio between velocities dynamically
flows to zero, which has been confirmed in the non-perturbative solution
in $2 \leq d < 3$\cite{2016arXiv160806927S,SDWgenD}. 
The $\ep$-expansion provides an independent justification
for the  ansatz used in the non-perturbative solution.

\section*{Acknowledgments}

We thank Shouvik Sur for initial collaboration on this project. 
This research was supported in part by the Natural Sciences and Engineering Research Council of Canada. 
Research at the Perimeter Institute is supported in part by the Government of Canada through Industry Canada, and by the Province of Ontario through the Ministry of Research and Information.


\newpage
\onecolumngrid
\appendix

\setcounter{figure}{0}
\setcounter{subfigure}{0}
\setcounter{table}{0}
\setcounter{equation}{0}

\renewcommand\thefigure{A\arabic{figure}}
\renewcommand\thetable{A\arabic{table}}
\renewcommand\theequation{A\arabic{equation}}

\section{The beta functions and the anomalous dimensions}

In this section we summarize the expressions 
for the beta functions and the anomalous dimensions
derived from the minimal subtraction scheme.
More details can be found in Ref. \cite{PhysRevB.91.125136}.
The renormalized action is given by
the sum of the classical action and the counter terms
which can be expressed in terms of bare fields and bare couplings,
\beqa
\mc{S}_{ren} \nn &=& 
\sum_{n=1}^4 
\sum_{\s=1}^{N_c} 
\sum_{j=1}^{N_f} 
\int dk_B ~
\bar{\Psi}_{B;n,\s,j}(k_B) 
\Bigl[ i  \mathbf{\Gamma} \cdot \mathbf{K}_B  
+ i \g_{d-1} \eps_n( \vec k_B; v_B ) \Bigr] 
\Psi_{B;n,\s,j}(k)
\\ \nn &&  
+ \f14 \int dq_B ~ 
\Bigl[ \abs{\mathbf{Q}_B}^2 + ~ c_B^2 \abs{\vec q_B}^2
\Bigr] \tr{ \Phi_B(-q_B)  ~ \Phi_B(q_B) } 
\\ \nn && 
+ i \frac{g_B}{\sqrt{N_f}}
\sum_{n=1}^4
\sum_{\s,\s'=1}^{N_c}
\sum_{j=1}^{N_f} 
\int dk_B \, dq_B ~  
\Bigl[
\bar{\Psi}_{B;\bar n,\s,j}(k_B + q_B) 
\Phi_{B;\s,\s'}(q_B)
\g_{d-1} 
\Psi_{B;n,\s',j} (k_B) 
\Bigr]  
\\ \nn &&  
+ \f14 \int dk_{1B} \, dk_{2B} \, dq_{B} ~
\Bigl[ 
u_{1B} \tr{ \Phi_B(k_{1B} + q_B)  \Phi_B(k_{2B}-q_B) } \tr { \Phi_B(-k_{1B})  \Phi_B(-k_{2B}) } 
\\ && 
+ u_{2B} \tr{ \Phi_B(k_{1B} + q_B)  \Phi_B(k_{2B}-q_B)   \Phi_B(-k_{1B})  \Phi_B(-k_{2B}) } 
\Bigr].
\label{eq: S_ren_bare}
\eeqa
The renormalized quantities are related to the bare ones through
$\mathbf{K}  = \mathcal{Z}_\tau^{-1} ~ \mathbf{K}_B$, 
$\vec k = \vec{k}_B$, 
$\Psi_{n,\sigma,j}(k) = \mathcal{Z}^{- \half}_\psi ~ \Psi_{B;n,\sigma,j}(k_B)$, 
$\Phi(q) = \mathcal{Z}^{- \half}_\phi ~ \Phi_B(q_B)$,  
$v = \frac{\mathcal{Z}_3}{\mathcal{Z}_2} ~ v_B$, 
$c =  \lt[\frac{\mathcal{Z}_\phi ~ \mathcal{Z}_\tau^{d-1}   }{\mathcal{Z}_5 }\rt]^{\half}  ~ c_B$, 
$g = \frac{\mathcal{Z}_\psi ~ \mathcal{Z}_\phi^\half  ~ \mathcal{Z}_{\tau}^{2(d-1)} }{\mc{Z}_6 } ~ \mu^{-\frac{3 - d}{2}} ~ g_B$,
$u_1 = \frac{\mathcal{Z}_\phi^2 \mathcal{Z}_{\tau}^{3(d-1)} }{\mathcal{Z}_7 } ~ \mu^{-(3 - d)} ~ u_{1B}$, 
$u_2 = \frac{\mathcal{Z}_\phi^2 \mathcal{Z}_{\tau}^{3(d-1)} }{\mathcal{Z}_8 } ~ \mu^{-(3 - d)} ~ u_{2B}$,
where $\mathcal{Z}_\tau = \frac{\mathcal{Z}_1}{\mathcal{Z}_3}$, 
$\mathcal{Z}_\psi = \mathcal{Z}_1 ~\mathcal{Z}_{\tau}^{-d}$,
$\mathcal{Z}_\phi = \mathcal{Z}_4 ~\mathcal{Z}_{\tau}^{-(d +1)}$
and $ \mathcal{Z}_n = 1 + \mathcal{A}_n $. 
The scaling dimension of $\vec k$ is fixed to be $1$.
By requiring that the bare quantities are independent of the scale $\mu$, 
we obtain 
the dynamical critical exponent,
the anomalous dimensions
and the beta functions as
\eqaln{
	z &= \lt[ 1 + 
	( Z^\prime_{1,1} - Z^\prime_{3,1} ) \rt]^{-1}, 
	\label{eq: z} \\
	\eta_\psi &=  - \f{\ep}{2}~ z  \lt( Z^\prime_{1,1} - Z^\prime_{3,1} \rt) + 
	\f12~ z  \lt( 2 Z^\prime_{1,1} - 3 Z^\prime_{3,1} \rt),  
	\label{eq: eta_psi} \\
	\eta_\phi &=  - \f{\ep}{2}~ z  \lt( Z^\prime_{1,1} - Z^\prime_{3,1} \rt) + 
	\f12 ~ z \lt( 4 Z^\prime_{1,1} - 4 Z^\prime_{3,1} - Z^\prime_{4,1} \rt), 
	\label{eq: eta_phi} \\
	\f{dv}{dl} &=    -z ~ v \lt( Z^\prime_{2,1} -  Z^\prime_{3,1} \rt), 
	\label{eq: beta_v} \\
	\f{dc}{dl} &=   -\f12~ z ~ c 
	\lt( 2Z^\prime_{1,1} -2 Z^\prime_{3,1} -  Z^\prime_{4,1} + Z^\prime_{5,1} \rt), 
	\label{eq: beta_c} \\
	\f{dg}{dl} &=   z ~g \lt[ \f{\ep}{2} + \f12 \lt( 2 Z^\prime_{3,1} +  
	Z^\prime_{4,1} - 2 Z^\prime_{6,1} \rt) \rt], 
	\label{eq: beta_g} \\
	\f{du_1}{dl} &=   z ~ u_1 
	\lt[ 
	\ep -  
	\lt(  2 Z^\prime_{1,1} - 2  Z^\prime_{3,1} - 2  Z^\prime_{4,1} + Z^\prime_{7,1} \rt) 
	\rt], 
	\label{eq: beta_u1} 
	\\
	%
	\f{du_2}{dl} &=   z ~ u_2 
	\lt[ 
	\ep -  
	\lt(  2 Z^\prime_{1,1} - 2  Z^\prime_{3,1} - 2  Z^\prime_{4,1} + Z^\prime_{8,1} \rt) 
	\rt],
	\label{eq: beta_u2} 
}
where $l = -\ln \mu$ is the logarithmic length scale, 
and $Z^\prime_{n,1} \equiv \lt( \f12~ g \partial_g + u_i \partial_{u_i} \rt) Z_{n,1}$.

\renewcommand\thefigure{B\arabic{figure}}
\renewcommand\thetable{B\arabic{table}}
\renewcommand\theequation{B\arabic{equation}}

\section{Computation of the boson self energy at two loops} 

In this section we compute the quantum corrections to the spatial part of the boson self-energy.
Among the two-loop diagrams, only Fig. \ref{fig:two-loop boson self energy} contributes.
It is written as
\beq
	\delta \Gamma_{0,2}^{2L} 
= -\frac{\mu ^{2 \epsilon }
	}{4} \f{4 g^4}{N_c N_f}
	\int dp ~
	\Upsilon_{0,2}^{2 L}(p) 
	\text{Tr}[\Phi (-p)
	\Phi (p)],
\eeq
where 
\beqa
	\Upsilon _{0,2}^{2 L}(p) &=&
	\sum _n
	\int
	dk \, dq ~
	\text{Tr}\left[\gamma _{d-1} G_n(q+k)
	\gamma _{d-1} G_{\bar{n}}(p+q+k)
	\gamma _{d-1} G_n(p+k) \gamma
	_{d-1} G_{\bar{n}}(k)\right] D(q).
	\label{eq:upsilon_02_a_defn}
\eeqa
Since we are interested in the momentum-dependent part, we set $\textbf{P} = 0$. 
We first perform the frequency integrations, which introduces four Feynman parameters $x_1,x_2,y_1,y_2$, followed by the spatial integrations. 
The final expression is given by
\beq
\Upsilon _{0,2}^{2 L;a}(\vec{p}) = 
-\f{1}{\ep} \f{h_5(v,c)}{v^2} (p_x^2 + p_y^2) + \mathcal{O}(\ep^0),
\eeq
where $h_5(v,c)$ is defined as
\beq
h_5(v,c) = -\f{2}{(4 \pi)^{2}} 
\int_0^1 dx_1  \int_{0}^{1-x_1} dx_2 \int_{0}^{1} dy_1 \int_{0}^{1-y_1} dy_2
\; \lt(A v^2 + B\rt),
\eeq
with
\beqa
\nn A &=& -\f{1}{128 \pi^2} \Bigg(
\frac{4 \left(b_1+b_2+b_3\right) \left(2
	\left(1-y_1-y_2\right)-\left(x_1+x_2\right) \left(1-2 y_1-2
	y_2\right)\right)}{\sqrt{a_1 \, a_2 \, a_3 \, a_4} \left(1-x_1-x_2\right)
	\left(x_1+x_2\right){}^2}
\\ \nn && -\frac{2 a_1 \left(b_1+b_2+b_3\right) \left(a_2 \, a_4 \, d_{2,3}+a_3 \,
	\left(a_4 \, d_{2,2}+a_2 \, d_{2,4}\right)\right)}{\left(a_1 \, a_2 \, a_3 \,
	a_4\right){}^{3/2}}
+ \frac{4 \left(d_{2,5}+d_{2,6}+d_{2,7}\right)}{\sqrt{a_1 \, a_2 \, a_3 \, a_4}}
\\ \nn && + \frac{3 \left(b_1+b_2+b_3\right) \left(1-y_1-y_2\right)
	d_{3,1}}{\sqrt{a_1 \, a_2 \, a_3} \, a_4^{5/2} \left(1-x_1-x_2\right){}^2
	\left(x_1+x_2\right){}^2}
+ \frac{a_1 \left(b_1+b_2+b_3\right) \left(1-y_1-y_2\right) \left(a_3 \,
	d_{3,2} + a_2 \, d_{3,3}\right)}{\left(a_1 \, a_2 \, a_3 \, a_4\right){}^{3/2}
	\left(1-x_1-x_2\right){}^2 \left(x_1+x_2\right){}^2}
\\ \nn && -\frac{2 a_1 \left(1-y_1-y_2\right) \left(a_2 \, a_4 \,
	\left(d_{3,7}+d_{3,8}+d_{3,9}\right)+a_3 \, \left(a_4 \,
	\left(d_{3,4}+d_{3,5}+d_{3,6}\right)+a_2 \,
	\left(d_{3,10}+d_{3,11}+d_{3,12}\right)\right)\right)}{\left(a_1 \, a_2
	\, a_3 \, a_4\right){}^{3/2} \left(1-x_1-x_2\right){}^2
	\left(x_1+x_2\right){}^2}
\Bigg),
\\ \nn B &=& A \text{ with } (b_2 \rightarrow -b_2, d_{2,6} \rightarrow -d_{2,6}, d_{3,5} \rightarrow -d_{3,5}, d_{3,8} \rightarrow -d_{3,8}, d_{3,11} \rightarrow -d_{3,11}).
\eeqa
Here $d_{n,m}$ are defined as
\beqa
\nn d_{2,2} &=& - c_1 \left(\left(1-x_1-x_2\right)
\left(x_1+x_2\right)\right){}^{-2}
\left(1-y_1-y_2\right),
\\ \nn d_{2,3} &=& \left(-1+x_1+x_2-c_2 \left(-1+c_4+x_1+x_2\right)+c_4 \left(2-x_1-x_2+c_1
\left(-1+c_4+x_1+x_2\right)\right)\right)
\\ \nn && \times \left(-1+y_1+y_2\right) \left(\left(1-x_1-x_2\right)\left(x_1+x_2\right)\right){}^{-2},
\\ \nn d_{2,4} &=&  
\Big(
c_1 c_5^2 \left(-1+y_1+y_2\right)+\left(-1+x_1+x_2\right) \left(-1+y_1+y_2\right)-c_8^2
\left(1-c_1 c_4^2-x_1-x_2+c_2 \left(-1+c_4+x_1+x_2\right)
\rt. \\ \nn && + \lt.
c_4 \left(-2+x_1+x_2-c_1
\left(-1+x_1+x_2\right)\right)\right) \left(-1+y_1+y_2\right)-c_5 \left(-1+x_1+x_2-c_8
\left(-2+c_2+x_1+x_2\right)
\rt. \\ \nn && + \lt.
c_1 \left(2-x_1-x_2+c_8 \left(-1+2 c_4+x_1+x_2\right)\right)\right)
\left(-1+y_1+y_2\right) + c_8 \left(-4+3 x_1+3 x_2+4 y_1-4 x_1 y_1+x_1^2 y_1
\rt. \\ \nn && - \lt.
4 x_2 y_1+2 x_1 x_2
y_1+x_2^2 y_1+4 y_2-4 x_1 y_2+x_1^2 y_2-4 x_2 y_2+2 x_1 x_2 y_2+x_2^2 y_2+c_2
\left(-2+x_1+x_2\right) \left(-1+y_1+y_2\right)
\rt. \\ \nn && - \lt.
c_4 \left(1-x_1-x_2+c_1 \left(-2+x_1+x_2\right)\right) \left(-1+y_1+y_2\right)\right)
\Big)
\left(\left(1-x_1-x_2\right)
\left(x_1+x_2\right)\right){}^{-2},
\eeqa
\beqa
\nn d_{2,5} &=&
\Big(
c_{11} \left(-c_9+c_8 c_{11}\right) \left(-1+x_1+x_2\right) \left(x_1+x_2\right)+\left(c_1 c_5^2
c_{11}^2+c_1 c_4^2 \left(1+c_9-c_8 c_{11}\right){}^2+c_9
\left(-1+x_1+x_2\right)
\rt. \\ \nn && - \lt.
\left(-1+c_2\right) c_9^2 \left(-1+x_1+x_2\right)-2 c_8 c_9 c_{11}
\left(-1+x_1+x_2\right)+c_5 c_9 c_{11} \left(2-x_1-x_2+c_1 \left(-1+x_1+x_2\right)\right)
\rt. \\ \nn && + \lt.
c_2 c_{11} \left(-1+c_8 c_{11}\right) \left(-2+c_5+x_1+x_2-c_8
\left(-1+x_1+x_2\right)\right)+c_{11} \left(-2-c_{11}+x_1+c_{11} x_1+x_2+c_{11} x_2
\rt. \rt. \\ \nn && - \lt. \lt. 
c_9 \left(-2+x_1+x_2\right){}^2+c_8^2 c_{11} \left(-1+x_1+x_2\right)+c_8 \left(1-x_1-x_2+c_{11}
\left(-2+x_1+x_2\right){}^2\right)\right)
\rt. \\ \nn && - \lt.
c_4 \left(1+c_9-c_8 c_{11}\right) \left(-1-c_{11}+2
c_1 c_{11}-2 c_1 c_5 c_{11}+2 c_8 c_{11}-c_1 c_8 c_{11}+c_2 \left(1+c_9-c_8
c_{11}\right)+c_{11} x_1-c_1 c_{11} x_1
\rt. \rt. \\ \nn && - \lt. \lt.
c_8 c_{11} x_1+c_1 c_8 c_{11} x_1+c_{11} x_2-c_1 c_{11}
x_2-c_8 c_{11} x_2+c_1 c_8 c_{11} x_2+c_9 \left(-2+x_1+x_2-c_1
\left(-1+x_1+x_2\right)\right)\right)
\rt. \\ \nn && - \lt. 
c_2 c_9 \left(-1+x_1+x_2+c_{11} \left(-2+c_5+x_1+x_2-2
c_8 \left(-1+x_1+x_2\right)\right)\right)-c_5 c_{11} \left(-1+c_{11} \left(-1+x_1+x_2
\rt. \rt. \rt. \\ \nn && - \lt. \lt. \lt.
c_8 \left(-2+x_1+x_2\right)+c_1 \left(2-x_1-x_2+c_8
\left(-1+x_1+x_2\right)\right)\right)\right)\right) \left(-1+y_1+y_2\right)
\Big)
\left(\left(1-x_1-x_2\right)
\left(x_1+x_2\right)\right){}^{-2},
\eeqa
\beqa
\nn d_{2,6} &=&
\left(\left(1-x_1-x_2\right)
\left(x_1+x_2\right)\right){}^{-2} 
\Big(
\left(-c_9 c_{12}+c_8 c_{11} \left(-1+2 c_{12}\right)\right) \left(-1+x_1+x_2\right)
\left(x_1+x_2\right)+\left(\left(\left(-1+c_2\right) c_8
\rt. \rt. \\ \nn && + \lt. \lt.
c_1 \left(c_5-c_4 c_8\right)\right)
\left(-\left(-1+c_5\right) c_{11}-c_4 \left(1+c_9-c_8 c_{11}\right)\right)
\left(-1+c_{12}\right)-\left(-1+c_5-c_4 c_8\right) \left(1+c_9-c_8 c_{11}
\rt. \rt. \\ \nn && - \lt. \lt.
c_2 \left(1+c_9-c_8 c_{11}\right)+c_1 \left(c_5 c_{11}+c_4 \left(1+c_9-c_8 c_{11}\right)\right)\right)
\left(-1+c_{12}\right)+\left(c_8 \left(1+2 c_{11}-c_5 c_{11}+c_1 c_5 c_{11}
\rt. \rt. \rt. \\ \nn && - \lt. \lt. \lt.
2 c_8 c_{11}+c_2 \left(-1+\left(-1+2 c_8\right) c_{11}\right)-\left(-1+c_1\right) c_4 \left(-1+\left(-1+2
c_8\right) c_{11}\right)\right)+c_9 \left(-1+\left(-1+c_1\right) c_5 \left(-1+c_{12}\right)
\rt. \rt. \rt. \\ \nn && + \lt. \lt. \lt.
2 \left(-1+c_2+c_4-c_1 c_4\right) c_8 \left(-1+c_{12}\right)+2 c_{12}-c_2 c_{12}-c_4 c_{12}+c_1
c_4 c_{12}\right)+\left(-1+c_8\right) \left(\left(-1+c_2-\left(-1+c_1\right) c_4\right)
c_{12}
\rt. \rt. \rt. \\ \nn && + \lt. \lt. \lt.
c_{11} \left(1+\left(-1+c_1\right) c_5-2 \left(1+\left(-1+c_1\right)
c_5+\left(-1+c_2+c_4-c_1 c_4\right) c_8\right) c_{12}\right)\right)\right)
\left(1-x_1-x_2\right)
\rt. \\ \nn && - \lt.
c_8 c_{11} \left(-1+c_{12}\right)
\left(-1+x_1+x_2\right){}^2+\left(c_9-c_8 c_{11}\right) c_{12}
\left(-1+x_1+x_2\right){}^2\right) \left(1-y_1-y_2\right)
\Big),
\eeqa
\beqa
\nn d_{2,7} &=&
\left(\left(1-x_1-x_2\right)
\left(x_1+x_2\right)\right){}^{-2} 
\Big(
\left(\left(-1+c_{12}\right) \left(c_1 \left(c_5-c_4 c_8\right) \left(1+c_5
\left(-1+c_{12}\right)-\left(-1+c_4\right) c_8 \left(-1+c_{12}\right)-2
c_{12}\right)
\rt. \rt. \\ \nn && - \lt. \lt.
\left(1+c_2 \left(-1+c_4\right)-2 c_4\right) c_8^2
\left(-1+c_{12}\right)+\left(-1+c_5\right) c_{12}+c_8 \left(-2+c_2 \left(1+c_5
\left(-1+c_{12}\right)-2 c_{12}\right)-2 c_5 \left(-1+c_{12}\right)
\rt. \rt. \rt. \\ \nn && + \lt. \lt. \lt.
4 c_{12}-c_4 c_{12}\right)\right)-c_8 x_1+\left(1-c_2+\left(-1+c_1\right) c_4\right) c_8^2
x_1+\left(-1+c_1\right) c_5 \left(-1+c_8\right) c_{12} x_1+c_8 c_{12} x_1-2 c_8^2 c_{12} x_1
\rt. \\ \nn && - \lt.
\left(-1+c_1\right) c_5 c_8 \left(-1+c_{12}\right) c_{12}
x_1+\left(1-c_2+\left(-1+c_1\right) c_4\right) c_8^2 c_{12}^2 x_1+c_{12}
\left(-1+\left(1+\left(-1+c_1\right) c_5\right) c_{12}\right) x_1
\rt. \\ \nn && + \lt.
c_8 \left(-1+c_{12}\right) x_2+\left(1-c_2+\left(-1+c_1\right) c_4\right) c_8^2 \left(-1+c_{12}\right){}^2
x_2+\left(1+\left(-1+c_1\right) c_5\right) \left(-1+c_{12}\right) c_{12} x_2
\rt. \\ \nn && + \lt.
c_8\left(\left(c_2-\left(-1+c_1\right) c_4\right) c_{12} \left(\left(-1+2 c_8+c_{12}\right)
x_1+\left(-1+c_{12}\right) x_2\right)-\left(-1+c_1\right) c_5
\left(x_1+\left(-1+c_{12}\right){}^2 x_2\right)\right)\right) \left(-1+y_1+y_2\right)
\\ \nn && +
c_8\left(-1+c_{12}\right) c_{12} \left(x_1+x_2\right) \left(3+\left(-4+x_1+x_2\right)
y_1+\left(-4+x_1+x_2\right) y_2\right)
\Big),
\eeqa
\beqa
\nn d_{3,1} &=& c_8 \left(1-c_5+c_4 c_8\right) \left(-\left(-1+c_2\right) c_8+c_1
\left(-c_5+c_4 c_8\right)\right),
\\ \nn d_{3,2} &=& c_1 c_8,
\\ \nn d_{3,3} &=& 1-c_5+c_2 \left(-1+c_5-3 c_4 c_8\right)+c_4 \left(3 c_8+c_1 \left(1-2
c_5+3 c_4 c_8\right)\right),
\\ \nn d_{3,4} &=& c_1 c_{11} \left(-c_9+c_8 c_{11}\right),
\\ \nn d_{3,5} &=& -c_1 \left(c_8 c_{11} \left(1-2 c_{12}\right)+c_9 c_{12}\right),
\\ \nn d_{3,6} &=& c_1 c_8 \left(-1+c_{12}\right) c_{12},
\\ \nn d_{3,7} &=& c_{11} \left(\left(-1+c_2\right) \left(-1+c_5\right) c_{11}+c_1 c_4^2 \left(-2-3 c_9+3 c_8
c_{11}\right)
\rt. \\ \nn && + \lt.
c_4 \left(-2-3 c_9+c_1 c_{11}-2 c_1 c_5 c_{11}+3 c_8 c_{11}+c_2 \left(2+3 c_9-3
c_8 c_{11}\right)\right)\right),
\eeqa
\beqa
\nn d_{3,8} &=& c_4 \left(-1+c_2-c_1 c_4\right) \left(2+3 c_9\right) c_{12}-\left(-1+c_5-3 c_4 c_8-c_1 c_4
\left(1-2 c_5+3 c_4 c_8\right)+c_2 \left(1-c_5+3 c_4 c_8\right)\right) c_{11} \left(-1+2
c_{12}\right),
\\ \nn d_{3,9} &=& -\left(-1+c_5-3 c_4 c_8-c_1 c_4 \left(1-2 c_5+3 c_4 c_8\right)+c_2 \left(1-c_5+3 c_4
c_8\right)\right) \left(-1+c_{12}\right) c_{12},
\\ \nn d_{3,10} &=& c_1 c_4^2 c_8+\left(1-c_5+c_2 \left(-1+c_5-3 c_4 c_8\right)+c_4 \left(3 c_8+c_1 \left(1-2 c_5+3
c_4 c_8\right)\right)\right) c_9^2+3 c_8 \left(-1+c_5-c_4 c_8\right) c_{11} \left(1-c_2
\rt. \\ \nn && + \lt.
2\left(c_1 c_5+\left(-1+c_2-c_1 c_4\right) c_8\right) c_{11}\right)+c_9
\left(\left(-1+c_2\right) \left(-1+c_5-2 c_4 c_8\right)-3 \left(\left(-1+c_2\right) c_8
\left(-2+2 c_5-3 c_4 c_8\right)
\rt. \rt. \\ \nn && + \lt. \lt.
c_1 \left(c_5^2+c_4 c_8 \left(2+3 c_4 c_8\right)-c_5 \left(1+4
c_4 c_8\right)\right)\right) c_{11}\right)-c_4 \left(c_1 \left(-1+2 c_5\right) c_9+c_8
\left(-1+c_2+2 \left(-1+c_2-2 c_1 c_4\right) c_9
\rt. \rt. \\ \nn && + \lt. \lt.
3 \left(-\left(-1+c_2\right) c_8+c_1 \left(1-2c_5+2 c_4 c_8\right)\right) c_{11}\right)\right),
\eeqa
\beqa
\nn d_{3,11} &=&
-c_1 \left(-1+c_5\right) c_5 c_9 \left(-2+3 c_{12}\right)+3 c_4 \left(1-c_2+c_1 c_4\right) c_8^3
c_{11} \left(-3+4 c_{12}\right)+c_8^2 \left(-c_1 c_4^2 \left(2+3 c_9\right) \left(-2+3
c_{12}\right)
\rt. \\ \nn && + \lt.
3 \left(-1+c_2\right) \left(-1+c_5\right) c_{11} \left(-3+4 c_{12}\right)+c_4
\left(4-9 c_1 c_{11}+18 c_1 c_5 c_{11}+c_9 \left(6-9 c_{12}\right)-6 c_{12}+12 c_1 c_{11}
c_{12}
\rt. \rt. \\ \nn && - \lt. \lt.
24 c_1 c_5 c_{11} c_{12}+c_2 \left(2+3 c_9\right) \left(-2+3
c_{12}\right)\right)\right)+c_8 \left(\left(-1+c_5\right) \left(1+2 c_9\right) \left(-2+3
c_{12}\right)-c_2 \left(-1+c_5\right) \left(1+2 c_9\right) \left(-2+3 c_{12}\right)
\rt. \\ \nn && + \lt.
c_1 \left(c_4 \left(-1+2 c_5\right) \left(1+2 c_9\right) \left(-2+3 c_{12}\right)+3
\left(-1+c_5\right) c_5 c_{11} \left(-3+4 c_{12}\right)\right)\right),
\\ \nn d_{3,12} &=& 3 c_8 \left(1-c_5+c_4 c_8\right) \left(-\left(-1+c_2\right) c_8+c_1 \left(-c_5+c_4
c_8\right)\right) \left(1-3 c_{12}+2 c_{12}^2\right),
\eeqa
where $a_n$, $b_n$ and $c_n$ are given by
\beqa
\nn a_{1} &=& -\mathcal{X}_1
	\left(-1+y_1+y_2\right)
	\lt(4 v^2 \left(-1+x_1+x_2\right)
	\left(x_1+x_2\right)\rt)^{-1},
\\ \nn a_{2} &=& - \mathcal{X}_2 \left(-1+y_1+y_2\right) \lt(\mathcal{X}_1 \left(-1+x_1+x_2\right) \left(x_1+x_2\right)\rt)^{-1},
\\ \nn a_{3} &=& \mathcal{X}_3 \lt(\left(x_1+x_2\right) \mathcal{X}_2\rt)^{-1},
\\ \nn a_{4} &=&	\mathcal{X}_4 \lt(\left(x_1+x_2\right) \mathcal{X}_3\rt)^{-1},
\\ \nn b_{1} &=&	
	\mathcal{X}_4^{-1} \,
	c^2 x_1 y_1 \left(-1+y_1+y_2\right) \left(c^2 \left(-1+x_1\right)
	x_1 y_2+\left(-1+c^2-v^2\right) x_2^2 y_2+x_2 \left(-c^2+c^2
	y_1+\left(-1+2 c^2-v^2\right) x_1 y_2\right)\right),
\\ \nn b_{2} &=&	
-\mathcal{X}_4^{-1} \, 2 c^2 \left(-1+v^2\right) x_1 x_2 \left(x_1+x_2\right) y_1 y_2
	\left(-1+y_1+y_2\right),
\\ \nn b_{3} &=& \mathcal{X}_4^{-1} \,
	c^2 x_2 \left(\left(-1+c^2-v^2\right) x_1^2 y_1+c^2
	\left(-1+x_2\right) x_2 y_1+x_1 \left(\left(-1+2 c^2-v^2\right) x_2
	y_1+c^2 \left(-1+y_2\right)\right)\right) y_2
	\left(-1+y_1+y_2\right),
\\ \nn c_1 &=& -\mathcal{X}_1^{-1} \, c^2 \left(-1+v^2\right)
	\left(-1+x_1+x_2\right), 
\\ \nn c_2 &=& -\mathcal{X}_1^{-1} \, 4 v^2 x_1,
\\ \nn c_4& =& c_1 \, \mathcal{X}_1 \, \mathcal{X}_2^{-1} \, x_1,
\\ \nn  c_5 &=& - \mathcal{X}_1 \, \mathcal{X}_2^{-1} \, x_2,
\\ \nn c_8 &=& \mathcal{X}_3^{-1} \, c^2 \left(-1+v^2\right) x_1 x_2 \left(-1+y_1+y_2\right),
\\ \nn c_9 &=& \mathcal{X}_3^{-1} \,
c^2 x_1 \left(c^2-c^2 x_1+\left(1-c^2+v^2\right) x_2\right)
	\left(-1+y_1+y_2\right),
\\ \nn c_{11} &=& -\mathcal{X}_4^{-1} \, \lt(c^2 \left(-1+v^2\right) x_1 x_2 \left(x_1+x_2\right) y_1
	\left(-1+y_1+y_2\right)\rt),
\\ \nn c_{12} &=&	-\mathcal{X}_4^{-1} \, \lt(c^2 x_2 \left(\left(1-c^2+v^2\right) x_1^2 y_1-c^2
	\left(-1+x_2\right) x_2 y_1+x_1 \left(\left(1-2 c^2+v^2\right) x_2
	y_1-c^2 \left(-1+y_2\right)\right)\right)
	\left(-1+y_1+y_2\right)\rt)
\eeqa
with
\beqa
\nn \mathcal{X}_1 &=& \left(c^2+\left(-4+c^2\right) v^2\right) x_1+c^2
\left(1+v^2\right) \left(-1+x_2\right),
\\ \nn \mathcal{X}_{2} &=&	
c^2 \left(-1+c^2-v^2\right) x_1^2+c^2 \left(-1+x_2\right)
\left(-c^2+\left(-1+c^2-v^2\right) x_2\right)
+
x_1 \left(c^2 \left(1-2 c^2+v^2\right)+2 \left(c^4+2 v^2-c^2
\left(1+v^2\right)\right) x_2\right),
\\ \nn \mathcal{X}_{3} &=&	
c^2 \left(-1+c^2-v^2\right) x_1^3 y_1+c^2 \left(-1+x_2\right) x_2
\left(-c^2+\left(-1+c^2-v^2\right) x_2\right) y_1+x_1 \left(\left(3
c^4+4 v^2-3 c^2 \left(1+v^2\right)\right) x_2^2 y_1
\rt. \\ \nn && \lt. +
c^2 x_2 \left(\left(1-3 c^2+v^2\right) y_1+\left(-1+c^2-v^2\right)
\left(-1+y_2\right)\right)-c^4 \left(-1+y_2\right)\right)+x_1^2
\left(\left(c^2 \left(1-c^2+v^2\right)
\rt. \rt. \\ \nn && \lt. \lt. +
\left(3 c^4+4 v^2-3 c^2
\left(1+v^2\right)\right) x_2\right) y_1+c^4
\left(-1+y_2\right)\right),
\\ \nn \mathcal{X}_4 &=& 
	c^2 \left(-1+c^2-v^2\right) x_1^4 y_1
	y_2+x_1^3 \left(\left(c^2 \left(1-c^2+v^2\right)+4
	\left(-1+c^2\right) \left(c^2-v^2\right) x_2\right) y_1+c^4
	\left(-1+y_2\right)\right) y_2
	\\ \nn && +
	c^2 \left(-1+x_2\right) x_2^2 y_1
	\left(-c^2+c^2 y_1+\left(-1+c^2-v^2\right) x_2 y_2\right)+x_1 x_2
	\left(c^2 \left(-1+2 c^2-v^2\right) x_2 y_1^2
	\rt. \\ \nn && \lt. +
	c^2 \left(-1+y_2\right) \left(-c^2+\left(-1+c^2-v^2\right) x_2
	y_2\right)+y_1 \left(4 \left(-1+c^2\right) \left(c^2-v^2\right)
	x_2^2 y_2+c^4 \left(-1+2 y_2\right)
	\rt. \rt. \\ \nn && \lt. \lt. +
	c^2 x_2 \left(1-2
	c^2+v^2+\left(1-3 c^2+v^2\right) y_2\right)\right)\right)+x_1^2
	\left(2 \left(3 c^4+4 v^2-3 c^2 \left(1+v^2\right)\right) x_2^2 y_1
	y_2-c^4 \left(-1+y_2\right) y_2 
	\rt. \\ \nn && \lt.  +
	c^2 x_2 \left(\left(-1+c^2-v^2\right) y_1^2+\left(-1+2 c^2-v^2\right)
	\left(-1+y_2\right) y_2+y_1 \left(1-c^2+v^2+\left(1-3 c^2+v^2\right)
	y_2\right)\right)\right).
\eeqa

\begin{figure}[h]
	\centering 
	\includegraphics[width=3.3in]{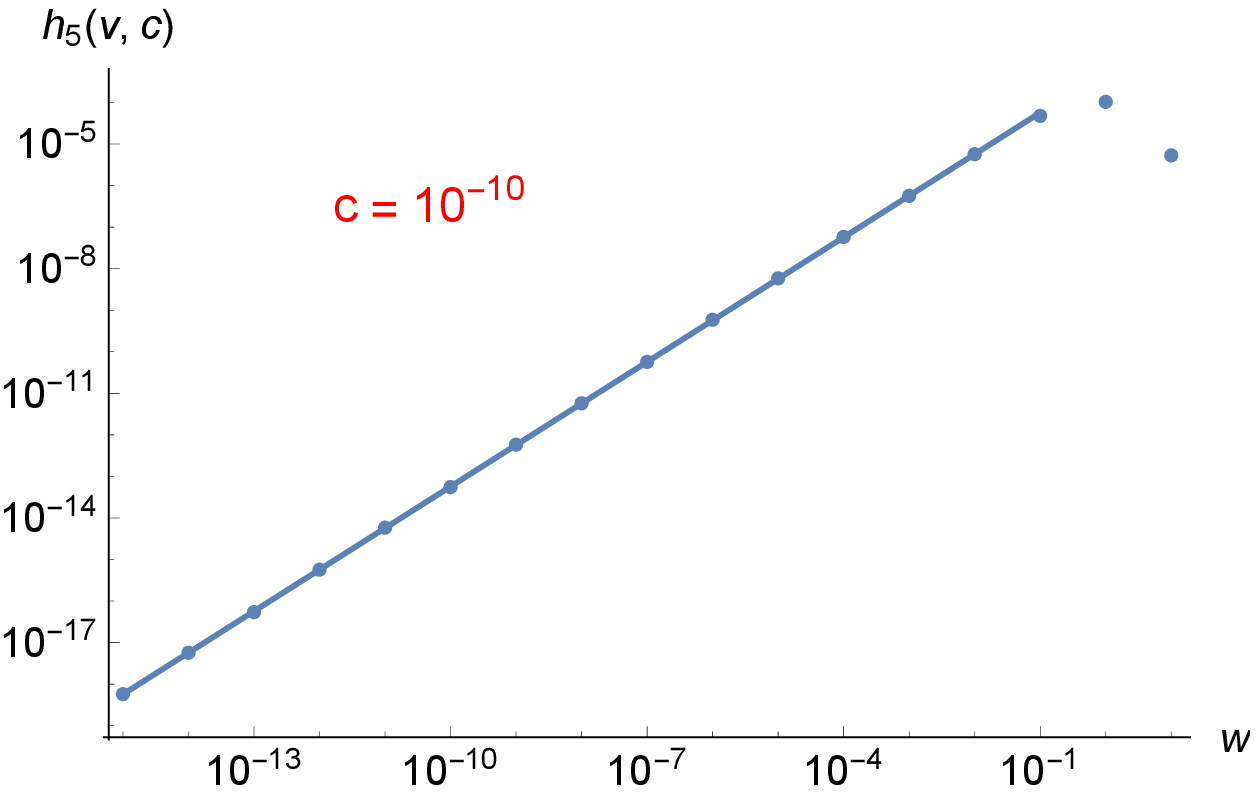}
	\caption{
	The dots represent $h_5(v,c)$ 
	evaluated as a function of $w=v/c$ at fixed $c = 10^{-10}$.
The line represents $L (w) = 5.7 \times 10^{-4} \, w$.
It is noted that $h_5(v,c)$ deviates from the line beyond $w \sim 0.1$.
}
	\label{fig:h_5 scaling}
\end{figure}

To extract the leading behavior of $h_5(v,c)$ in the limit $v, c, w \equiv v/c$ are small, we approximate the integrand in $h_5(v,c)$ by its leading order term in this limit. This gives $h_5 (v,c) = h_5^* \, w$ with $h_5^* \approx 5.7 \times 10^{-4}$ to the leading order in $v,c, w$. 
In Fig. (\ref{fig:h_5 scaling}), we show the full $h_5(v,c)$ as a function of $w = \f{v}{c}$ for a small value of $c$, 
which confirms the linear behavior in the small $w$ limit.
The two-loop contribution to the quantum effective action is
\beq
	\delta \Gamma_{0,2}^{2L} = \f{1}{\ep} \f{4}{N_c N_f} \f{g^4}{v^2 c^2} h_5(v,c)
	\int 
	dp ~ 
	\f14 c^2 |\vec{p}|^2 \text{Tr}[\Phi (-p)\Phi (p)] ~ + ~ \mathcal{O}(\ep^0).
\label{eq:2LG}
\eeq
Combining \eq{eq:2LG} with the one-loop quantum effective action obtained in Ref. \cite{PhysRevB.91.125136},
we obtain the counter terms as
\beqa
	Z_{1,1} &=&  - \frac{(N_c^2 - 1)}{4 \pi^2 N_c N_f } ~ 
	\frac{g^2}{c} ~ h_1(v,c), 
	\\
	Z_{2,1} &=&   \frac{(N_c^2 - 1)}{4 \pi^2 N_c N_f } ~ 
	\frac{g^2}{c} ~ h_2(v,c), 
	\\
	Z_{3,1} &=&  - Z_{2,1}, 
	\\
	Z_{4,1} &=&  - \frac{1}{4 \pi} ~ \frac{g^2}{v}, 
	\\
	Z_{5,1} &=& - \f{4}{N_c N_f} \f{g^4}{v^2 c^2} h_5(v, c), 
	\\
	Z_{6,1} &=&  - \frac{1}{8\pi^3 N_c N_f} 
	~ \frac{g^2}{c} ~ h_3(v,c), 
	\\
	Z_{7,1} &=& \frac{1}{2 \pi^2 c^2} \lt[
	(N_c^2+7)u_1 
	+ 2 \lt( 2 N_c - \frac{3}{N_c} \rt) u_2 
	+ 3 \lt( 1 + \frac{3}{N_c^2} \rt) \frac{u_2^2}{u_1}  
	\rt], 
	\\
	Z_{8,1} &=& \frac{1}{2 \pi^2 c^2} \lt[
	12 u_1  
	+ 2 \lt( N_c - \frac{9}{N_c} \rt) u_2  
	\rt].
	\label{eq: modified one loop Z_n}
\eeqa
Here $h_1(v,c)$, $h_2(v,c)$, and $h_3(v,c)$ are given by\cite{PhysRevB.91.125136}
\beqa
h_1(v,c) &=& \int_{0}^{1} dx \, \sqrt{\frac{1-x}{c^2 + (1+v^2-c^2)x}} 
= \frac{\pi  \left(1+v^2\right)-2 c \sqrt{1-c^2+v^2}-i \left(1+v^2\right)
	\left(\log \left(1+v^2\right)-2 \log \left(i
	c+\sqrt{1-c^2+v^2}\right)\right)}{2 \left(1-c^2+v^2\right)^{3/2}},
\nn \\ h_2(v,c) &=& c^2 \int_{0}^{1} dx \, \sqrt{\frac{1-x}{(c^2 + (1+v^2-c^2)x)^3}}
= -\frac{c \left(-2 \sqrt{1-c^2+v^2}-i c \left(\log \left(-1-v^2\right)-2
	\log \left(i
	c+\sqrt{1-c^2+v^2}\right)\right)\right)}{\left(1-c^2+v^2\right)^{3/2
	}},
\nn \\ h_3(v,c) &=& \int_{0}^{1} dx_1 \int_{0}^{1-x_1} dx_2 \frac{\pi  c \left(8 v^2 x_1 x_2+2 c^4 \left(-1+x_1+x_2\right){}^2-c^2
	\left(-1+x_1+x_2\right) \left(1+2 x_1+2 x_2+v^2 \left(-1+2 x_1+2
	x_2\right)\right)\right)}{\left(4 v^2 x_1 x_2+c^4
	\left(-1+x_1+x_2\right){}^2-c^2 \left(1+v^2\right)
	\left(-1+x_1+x_2\right) \left(x_1+x_2\right)\right){}^{3/2}}. 
\nn
\eeqa
From the expressions for $Z_{n,1}$, we obtain the beta functions 
for $\l \equiv \f{g^2}{v}, x \equiv \f{g^2}{c^3}, w \equiv \f{v}{c}, \k_i \equiv \f{u_i}{c^2}$,
\beqa
	\f{d\l}{dl} &=&   z ~ \l \lt( \ep - \frac{\l}{4 \pi} + \frac{1}{4\pi^3 N_c N_f} 
	~ \l w ~ h_3(v,c) \rt),
	\label{eq:lambda beta function full}
	\\
	\f{d x}{dl} &=&   z ~ x \lt( \ep - 
	\frac{3 (N_c^2 - 1)}{4 \pi^2 N_c N_f } ~ 
	\l w ~ h_1(v,c) + \frac{(N_c^2 - 1)}{4 \pi^2 N_c N_f } ~ 
	\l w ~ h_2(v,c) + \frac{\l}{8 \pi} - \f{12}{N_c N_f} \f{\l x}{w} h_5(v, c) + \frac{\l w ~ h_3(v,c)}{4\pi^3 N_c N_f} 
	\rt),
	\label{eq:x beta function full}
	\\
	\f{dw}{dl} &=&  \f12 ~ z ~ w \lt( - \frac{(N_c^2 - 1)}{2 \pi^2 N_c N_f } ~ 
	\l w ~ h_1(v,c) - \frac{(N_c^2 - 1)}{2 \pi^2 N_c N_f } ~ 
	\l w ~ h_2(v,c) + \frac{\l}{4 \pi} - \f{8}{N_c N_f} \f{\l x}{w} h_5(v, c) \rt),
	\label{eq:w beta function full}
	\\
	\f{d\k_1}{dl} &=&   z ~ \k_1 \lt( \ep - 
	\frac{\l}{4 \pi} - \f{8}{N_c N_f} \f{\l x}{w} h_5(v, c) - \frac{1}{2 \pi^2} 
	\lt( (N_c^2+7)\k_1 
	+ 2 \lt( 2 N_c - \frac{3}{N_c} \rt) \k_2 
	+ 3 \lt( 1 + \frac{3}{N_c^2} \rt) \frac{\k_2^2}{\k_1} \rt)
	\rt),
	\label{eq:kappa_1 beta function full}
	\\
	\f{d\k_2}{dl} &=&   z ~ \k_2 \lt( \ep - 
	\frac{\l}{4 \pi} - \f{8}{N_c N_f} \f{\l x}{w} h_5(v, c) - 
	\frac{1}{2 \pi^2} \lt( 12 \k_1 + 2 \lt( N_c - \frac{9}{N_c} \rt) \k_2  \rt)
	\rt).
	\label{eq:kappa_2 beta function full}
\eeqa
The leading order behavior of $h_i(v,c)$ in the limit of small $v,c, w$ are 
$h_1(v,c) = \f{\pi}{2}$,
$ h_2(v,c) = 2 c$,
$ h_3(v,c) = 2 \pi^2$.

\renewcommand\thefigure{C\arabic{figure}}
\renewcommand\thetable{C\arabic{table}}
\renewcommand\theequation{C\arabic{equation}}

\section{Upper bound of higher-loop diagrams}

Here we estimate the magnitude of higher-loop diagrams 
without self-energy insertions at the M1L fixed point.
Since $\kappa_i=0$ at the fixed point, 
we consider diagrams made of Yukawa vertices only.
The discussion closely follows Appendix A of Ref. \cite{2016arXiv160806927S},
and we will be brief here.
A general $L$-loop diagram can be written as
\beq
	I \sim g^{V}  \int \prod_{r = 1}^{L} dp_r
	\lt(\prod_{l = 1}^{I_f} \f{1}{\textbf{K}_l \cdot \boldsymbol{\G} + \eps_{n_l}(k_l) \g_{d-1}} \rt)
	\lt(\prod_{m = 1}^{I_b} \f{1}{|\textbf{Q}_m|^{2} + c^{2} (q_{m,x}^{2} + q_{m,y}^{2})} \rt).
	\label{eq:L-loop integral}
\eeq
Here $V$ is the number of Yukawa vertices.
$I_f, I_b$ are the number of fermion and boson propagators, respectively. 
$p_r$'s represent the internal momenta. 
$k_l$ ($q_m$) is the momentum that flows through the $l$-th fermion ($m$-th boson) propagator,
which is given by  a linear combination of the internal and external momenta. 
$n_l$ is the patch index of the $l$-th fermion line. 
Without loss of generality, 
we can focus on diagrams that involve only patches $1$ and $3$. 
We are ignoring the $\g$ matrices coming from the Yukawa vertices as they play no role in the estimation.

The dependence of $k_l$ and $q_m$ on the internal momenta 
is determined by the choice of loops.
One can choose the loop momenta such that 
$L-L_f$ boson propagators become exclusive propagators, 
in the sense that each of them depends exclusively on only one internal momentum,
where $L_f$ is the number of fermion loops.
Since the limit of small $v,c,w$ does not affect the frequency integrations, 
we focus on the spatial parts of the propagators.
The integrations for $p_{r,x}, p_{r,y}$ in \eq{eq:L-loop integral} can be written as
\beq
	\nn  I \sim g^{V} \int \displaystyle\prod_{r = 1}^{L} dp_{r,x} dp_{r,y}
	\lt(\displaystyle\prod_{m = 1}^{L - L_f} \f{1}{(cp_{m,x})^{2} + (cp_{m,y})^{2}} \rt)
	\lt(\displaystyle\prod_{l = 1}^{I_f} \f{1}{E_l(p)} \rt) R[p].
	\label{eq:L-loop integral 2}
\eeq
Here we have dropped the frequency variables and all the $\gamma$ matrices. 
The first group represents the exclusive boson propagators for the $L-L_f$ non-fermion loops. 
The second group represents all fermion propagators, 
and the energy of the fermion is written $E_l(p) \equiv \eps_l( k_{l}(p) )$,
where $k_l(p)$ is the momentum that flows through the $l$-th fermion propagator
which is a function of internal momenta.
$R[p]$ represents the remaining boson propagators in the diagram.

Now we change variables in a way that the divergence in the small $v,c$ limit becomes manifest. 
The first $L - L_f$ variables are chosen to be $p^{'}_{i} \equiv c p_{i,x} $ with $1 < i \leq L - L_f$. 
The remaining $L + L_f$ variables are chosen among $\{E_l(p)\}$. 
$\{p'_{i}, E_l(p)\}$ are expressed in terms of $\{v p_{r,x}, p_{r,y}\}$ as
\beqa
\left(
\begin{array}{c}
	p^{'}_{1} \\
	p^{'}_{2} \\
	\vdots \\
	p^{'}_{L - L_f} \\
	\\
	E_{1} \\
	E_{2} \\
	\vdots \\
	E_{I_f}
\end{array}
\right)
= 
\left(
\begin{array}{cc}
	\frac{c}{v} {\mathbb I}_{L - L_f} & 0 \\ 
	{\mathbb A} & {\mathbb V}
\end{array}
\right)
\left(
\begin{array}{c}
	v p_{1,x} \\
	v p_{2,x} \\
	\vdots \\
	v p_{L - L_f,x} \\
	\\
	v p_{L - L_f + 1,x} \\
	v p_{L - L_f + 2,x} \\
	\vdots \\
	v p_{L,x} \\
	p_{1,y} \\
	p_{2,y} \\
	\vdots \\
	p_{L,y} \\
\end{array}
\right).
\label{eq:change of coordinates matrix 1 Yukawa only}
\eeqa
Here, ${\mathbb I}_{a}$ is the $a \times a$ identity matrix.
$\mathbb{A}$ is an $I_f \times (L - L_f)$ matrix whose matrix elements are given by
$\mathbb{A}_{n,i} = \f{1}{v} \f{\partial E_n}{\partial p_{i,x}}$ with 
$1 \leq n \leq I_f$ and $1 \leq i \leq L - L_f$. 
$\mathbb{V}$ is an $I_f \times (L + L_f)$ matrix whose first $L_f$ columns are given by 
$\mathbb{V}_{n,a - (L-L_f)} = \f1v \f{\partial E_n}{\partial p_{a,x}}$ for 
$L - L_f + 1 \leq a \leq L$ while the remaining $L$ columns are given by 
$\mathbb{V}_{n, b + L_f} = \f{\partial E_n}{\partial p_{b,y}}$ for 
$1 \leq b \leq L$. 
In Ref. \cite{2016arXiv160806927S} it is shown that the $L + L_f$ column vectors of $\mathbb{V}$ are linearly independent. Therefore, there exist $L + L_f$ row vectors of $\mathbb{V}$ that are linearly independent, 
which we label to be the $l_k$-th rows with $k = 1, ... , (L + L_f)$. 
Let $\tilde{\mathbb{V}}$ be the $(L + L_f) \times (L + L_f)$ matrix consisting of these rows. 
Then we 
define $p^{'}_{L - L_f + k} \equiv E_{l_k}$ with $k = 1, ... , (L + L_f)$ 
as the remaining $(L + L_f)$ integration variables. 
The new momentum variables are given in terms of the old variables by
\beqa
\left(
\begin{array}{c}
	p^{'}_{1} \\
	p^{'}_{2} \\
	\vdots \\
	p^{'}_{2L}
\end{array}
\right)
= 
\left(
\begin{array}{cc}
	\frac{c}{v} {\mathbb I}_{L - L_f} & 0  \\
	\tilde{{\mathbb A}} & \tilde{{\mathbb V}}
\end{array}
\right)
\left(
\begin{array}{c}
	v p_{1,x} \\
	v p_{2,x} \\
	\vdots \\
	v p_{L - L_f,x} \\
	\\
	v p_{L - L_f + 1,x} \\
	v p_{L - L_f + 2,x} \\
	\vdots \\
	v p_{L,x} \\
	p_{1,y} \\
	p_{2,y} \\
	\vdots \\
	p_{L,y} \\
\end{array}
\right),
\eeqa
where $\tilde{\mathbb{A}}$ is the collection of the $l_k$-th rows of $\mathbb{A}$, with $k = 1, ... , (L + L_f)$. 
The Jacobian of this change of variables is given by $Y ^{-1} c^{-(L - L_f)} v^{-L_f}$, 
where $Y = |\det \tilde{\mathbb{V}}|$ is a numerical constant independent of $v,c$.
$Y$ is nonzero because $\tilde{\mathbb{V}}$ is invertible. 
In the new basis, it is manifest that 
for every integration variable $p_r^{'}$,
there is one propagator 
that guarantees the integrand decays at least as $1/p_r^{'}$, 
in the limit $c,v  \rightarrow 0$.
Since there is no sub-diagram with a positive degree of UV divergence,
the integrations over $p_r^{'}$ are at most logarithmically divergent
in the UV cut-off or $v,c$.
Therefore, the diagram is bounded by
\beqa
I \sim \f{g^{V}}{v^{L_f}c^{L - L_f}}, 
\label{eq:scaling v c Yukawa only_app}
\eeqa
up to potential logarithmic corrections in $v$ and $c$.

\renewcommand\thefigure{D\arabic{figure}}
\renewcommand\thetable{D\arabic{table}}
\renewcommand\theequation{D\arabic{equation}}

\section{Beyond the modified one-loop order}

In this appendix, 
we consider the effects of higher-loop diagrams 
in the small $w$ limit.
To the leading order in $w$,
the higher-loop diagrams that need to be considered
are the M1L diagrams 
in which the boson propagator is dressed with
the self-energy insertions
in  Figs. \ref{fig:one-loop diagrams}(a) 
and \ref{fig:two-loop boson self energy}.
An  insertion of the self-energy in Fig. \ref{fig:one-loop diagrams}(a) adds 
one power of $\l$ to $Z_{n,1}$,
while an insertion of the self-energy in Fig. \ref{fig:two-loop boson self energy} adds 
one power of $\l x $, up to logarithmic corrections in $c,v$ for both insertions. 
We write the general form of the counter terms from the higher-loop diagrams as
\beqa
Z_{1,1} &=& \l w \sum_{n, m=0}^{\infty} \l^{n+m} x^m a_{n,m}(c,v),
\label{eq:higher loop Z_1}
\\ Z_{2,1} &=& \f{(\l w)^{\f32}}{x^{\f12}} \sum_{n, m=0}^{\infty} \l^{n+m} x^m b_{n,m}(c,v),
\label{eq:higher loop Z_2}
\\ Z_{3,1} &=& - Z_{2,1},
\label{eq:higher loop Z_3}
\\ Z_{4,1} &=& - \f{1}{4 \pi} \l,
\label{eq:higher loop Z_4}
\\ Z_{5,1} &=& \l x \sum_{n, m=0}^{\infty} \l^{n+m} x^m h_{n,m}(c,v),
\label{eq:higher loop Z_5}
\\ Z_{6,1} &=& \l w \sum_{n, m=0}^{\infty} \l^{n+m} x^m r_{n,m}(c,v).
\label{eq:higher loop Z_6}
\eeqa
Here, $a_{n,m}(c,v), b_{n,m}(c,v), h_{n,m}(c,v), r_{n,m}(c,v)$ are functions 
that grow at most logarithmically in $c,v$.
For $n = m = 0$, they are independent of $c,v$, 
and given by 
$a_{0,0}(c,v) = - \f{(N_c^2 - 1)}{8 \pi N_c N_f }$, $b_{0,0}(c,v) = \f{(N_c^2 - 1)}{2\pi^2 N_c N_f }$, $h_{0,0}(c,v) = - \f{4 \, h_5^*}{N_c N_f}$, $r_{0,0}(c,v) = - \f{1}{4\pi N_c N_f}$. 
The relation $Z_{2,1} = - Z_{3,1}$ still holds because 
the external momentum can be passed through a single fermion line 
with the opposite patch index to the external lines.

We first establish that the fixed point still
exists in the presence of general logarithmic corrections in $v,c$.
It is straightforward to check that $w^*=0$ remains as a fixed point.
At $w=0$, the beta functions for $\lambda, x$ read
\beqa 
&&	\f{d\l}{dl} =  z \, \l \Bigg( \ep - \f{1}{4 \pi} \l 
	\Bigg), \\ 
	\label{eq:lambda beta function higher loop0}
&&		\f{d {x}}{dl} = z x \Bigg(
	\ep 
	+ \frac{1}{8 \pi} \l + \f32 \l x  \sum_{n, m=0}^{\infty} (n+2m+2) \, \l^{n+m} x^m 
	h_{n,m}(c,v)  
	\Biggr).
\label{eq:x beta function higher loop0}
\eeqa
While $\lambda$ still flows to $\l^* = 4 \pi \ep$,
$x$ no longer flows to an $\mathcal{O}(1)$ fixed point 
if $h_{n,m}(c,v)$  diverge logarithmically
in the small $v,c$ limit.
This may be regarded as an indication that the theory has an instability.
However, we show that such a runaway flow is an artifact of looking 
at the wrong parameter $x$ for general $\ep$.
In other words, the relative rate at which $v,c$ flow to zero depends on $\ep$, 
and we have to take the $\ep$-dependence into account in
choosing the variable that represents the fixed point.
To see this, we define a new variable $\tilde{x} \equiv \frac{x}{F(c,v) }$ with
\beqa
F(c,v) 
= 1 + \sum_{p = 1}^{\infty}  \ep^p \, f_{p}(c,v),
\eeqa
where  we leave open the possibility that 
$f_{p}(c,v)$ depends on both $c,v$ for the sake of full generality. 
The beta function for $\tilde{x}$ is given by
\beqa 
\nn \f{d \tilde{x}}{dl} =  z ~ \tilde{x} && \lt[
\ep + \lt(3 + \f{\partial_{\log (c)} F}{F} \rt) Z^\prime_{1,1} + \f{\partial_{\log (v)} F}{F} Z^\prime_{2,1}
- \lt(1 + \f{\partial_{\log (c)} F}{F} + \f{\partial_{\log (v)} F}{F}\rt) Z^\prime_{3,1} 
\rt. \\ && \lt. - \f12 \lt(1 + \f{\partial_{\log (c)} F}{F} \rt) Z^\prime_{4,1} 
+ \f32 \lt(1 + \f{\partial_{\log (c)} F}{3 F} \rt) Z^\prime_{5,1} - 2 Z^\prime_{6,1} 
\rt],
\label{eq:tx}
\eeqa
where $Z^\prime_{n,1} \equiv \lt(\f12 g \partial_g + u_i \partial_{u_i} \rt) Z_{n,1}$. 
The point of introducing $\tilde x$ is that
we can determine $F(c,v)$ such that $\tilde x$ flows to an $\mathcal{O}(1)$ fixed point, $\tilde{x}^*$. 
The conditions, $\f{d \l}{dl} = 0$ and $\f{d \tilde x}{dl} = 0$ imply
\beqa
&& 1 + 4\pi \,  \tilde{x}^* \lt(1 + \sum_{p = 1}^{\infty} \ep^p \, f_{p}(c,v) \rt) \sum_{n, m=0}^{\infty} \, (n + 2m + 2) \, \ep^{n+m} (\tilde{x}^*)^m \lt(1 + \sum_{p = 1}^{\infty} \ep^p \, f_{p}(c,v) \rt)^m
\tilde{h}_{n,m}(c,v) = 0
\label{eq:fixed point value self consistent equation for x higher epsilon solution plugged in}
\eeqa
to the leading order in $w$, 
where $\tilde{h}_{n,m}(c,v)= (4\pi)^{n+m} h_{n,m}(c,v)$. 
Eq. (\ref{eq:fixed point value self consistent equation for x higher epsilon solution plugged in})
can be solved for $\tilde x^*$ and $f_{p}(c,v)$ at every order in $\ep$.
For $\a =0$, we have
\beqa
\nn 1+8 \pi \tilde{x}^* \tilde{h}_{0,0} = 0 
\eeqa
which gives $ \tilde{x}^* = -\frac{1}{8 \pi  \tilde{h}_{0,0}} = \f{N_c N_f}{32 \pi \, h_5^*} = x^*$.
The equation for general $\alpha > 0$ contains only
$f_{\alpha^\prime}$ with $\a^{\prime} \leq \a$, from which $f_{\a}$
is uniquely fixed.
For example, the first few equations in the series  read
\beqa
2 \tilde{h}_{0,0}f_1 +4 \tilde{x}^* \tilde{h}_{0,1}+3 \tilde{h}_{1,0} = 0 && ~~\mbox{for ~$\alpha=1$}, \nn \\
2 \tilde{h}_{0,0}f_2 +6 (\tilde{x}^*)^2 \tilde{h}_{0,2}+f_1 \left(8 \tilde{x}^*
\tilde{h}_{0,1}+3 \tilde{h}_{1,0}\right)+5 \tilde{x}^* \tilde{h}_{1,1}+4
\tilde{h}_{2,0} = 0 && ~~\mbox{for ~$\alpha=2$}, \nn \\
\nn 2 \tilde{h}_{0,0}f_3 +4 f_1^2 \tilde{x}^* \tilde{h}_{0,1}+8 f_2 \tilde{x}^*
\tilde{h}_{0,1}+8 (\tilde{x}^*)^3 \tilde{h}_{0,3}+3 f_2 \tilde{h}_{1,0}+7 (\tilde{x}^*)^2
\tilde{h}_{1,2}
\\ \nn +2 f_1 \left(9 (\tilde{x}^*)^2 \tilde{h}_{0,2}+5 \tilde{x}^*
\tilde{h}_{1,1}+2 \tilde{h}_{2,0}\right)+6 \tilde{x}^* \tilde{h}_{2,1}+5
\tilde{h}_{3,0} = 0 && ~~\mbox{for ~$\alpha=3$}, 
\label{eq:constraints}
\eeqa
each of which fixes 
$f_{1}(c,v)$, $f_{2}(c,v)$, $f_{3}(c,v)$, respectively. 
Therefore,  $f_p(c,v)$ can be determined such that 
$\tilde x$ flows to an $\mathcal{O}(1)$ value to all orders in $\ep$.
At the fixed point with $(\l^*,\tilde{x}^*,w^*) = \left( 4 \pi \ep,  \f{N_c N_f}{32 \pi \, h_5^*}, 0 \right)$,
Eq. (\ref{eq:tx}) implies that $Z^\prime_{5,1} = -\ep$,
and $Z_{1,1}$, $Z_{2,1}$, $Z_{3,1}$, $Z_{6,1}$ 
in Eqs. (\ref{eq:higher loop Z_1}), (\ref{eq:higher loop Z_3}), (\ref{eq:higher loop Z_3}), (\ref{eq:higher loop Z_6}) vanish 
because $x$ is divergent at most logarithmically in $w$.
The same conclusion holds for all other higher-loop diagrams suppressed by $w$. 
As a result, the $\ep$-expansion is well defined, 
and the fixed point with $w^*=0$ persists to all orders in $\ep$.
Furthermore,  the critical exponents 
in Eqs. (\ref{eq: z}), (\ref{eq: eta_psi}), (\ref{eq: eta_phi})
do not receive perturbative corrections beyond the M1L order
at the fixed point.
This is a rather remarkable feature attributed to $w^*=0$.

The remaining question is whether 
the non-trivial fixed point remains attractive
to all orders in $\ep$. In the small $\ep$ limit this is indeed the case. 
For general $\ep$, we cannot prove this 
from the present perturbative expansion 
without actually computing the counter terms  to all orders in $\ep$.
However, from the non-perturbative calculation\cite{2016arXiv160806927S,SDWgenD},
it is shown that $w$ indeed flows to zero 
for any $0 < \ep \leq 1$.

\renewcommand\thefigure{E\arabic{figure}}
\renewcommand\thetable{E\arabic{table}}
\renewcommand\theequation{E\arabic{equation}}

\section{Computation of physical properties}

Here we provide some details of the derivation of the scaling forms of the Green's functions.
The fermion Green's function satisfies the renormalization group equation\cite{PhysRevB.91.125136},
\begin{align}\label{eq:Eqg}
\left[z\mathbf{K}\cdot\frac{\partial}{\partial\mathbf{K}}+\vec{k}\cdot\frac{\partial}{\partial \vec{k}} - \beta_w\frac{\partial}{\partial w} - \beta_x\frac{\partial}{\partial x} - \beta_\lambda\frac{\partial}{\partial \lambda}-(2\eta_\psi+z(d-1)-d)\right]G_n(k;w,x,\lambda)=0,
\end{align}
where we have set $\k_i = 0$ and $\beta_{\k_i} = 0$.
The solution to this equation is given by
\begin{align}\label{eq:SolG}
G_n(k;w_0,x_0,\lambda_0)&=\exp\left(-\mathfrak{I}_{\psi}(l) \right)
G_n(e^{l}\mathbf{K},e^{\mathfrak{I}_{z}}\vec{k};,w(l),x(l),\lambda(l)),
\end{align}
where 
\begin{align}
\mathfrak{I}_z(l) &= \int_0\limits^{l}\frac{\dd\ell}{z(\ell)},\label{eq:IZ}\\
\mathfrak{I}_{\psi}(l) &=\int_0\limits^{l}\dd\ell \left(\frac{2\eta_\psi(\ell)+z(\ell)(2-\epsilon)-(3-\epsilon)}{z(\ell)}\right)\label{eq:IPsi},
\end{align}
and $w(l), x(l), \l(l)$ are solutions to
$\frac{\dd w(l)}{\dd l} = - \frac{\beta_{w}}{z(l)}$,
$\frac{\dd \lambda(l)}{\dd l} = - \frac{\beta_{\lambda}}{z(l)}$,
$\frac{\dd x(l)}{\dd l} = - \frac{\beta_{x}}{z(l)}$
with initial conditions, $w(0)=w_0, \l(0)=\l_0, x(0)=x_0$.
Because all three parameters flow,
the full crossover structure is rather complicated.
However, $w$ decays at the slowest rate,
\begin{align}
w(l) &\stackrel{l\gg 1}{=} \frac{N_c N_f}{2^\frac{11}{3}(h^{*}_5)^\frac{1}{3}(N^2_c-1)^\frac{2}{3}}\frac{1}{\epsilon }\frac{1}{l^\frac{2}{3}},
\label{eq:wl}
\end{align}
and the crossover at low energies is dominated by the flow of $w$.
To the leading order in $w$ and $\epsilon$,
\begin{align}	
z - 1 &= \frac{(N^2_c-1)\epsilon}{2N_c N_f} w-\frac{32\sqrt{2}\sqrt{h^{*}_5}(N^2_c-1)\epsilon^\frac{3}{2}}{N^\frac{3}{2}_cN^\frac{3}{2}_f} w^\frac{3}{2},\\
\eta_{\psi}&=-\frac{(N^2_c-1)\epsilon(2-\epsilon)}{4N_c N_f} w + \frac{8\sqrt{2}\sqrt{h^{*}_5}(N^2_c-1)(5-2\epsilon)\epsilon^\frac{3}{2}}{N^\frac{3}{2}_cN^\frac{3}{2}_f} w^\frac{3}{2}.
\end{align}
Although $w$ flows to zero in the low energy limit,
the slow decay of $w$ renormalizes
the scaling of the frequency and the field at intermediate energy scales as  
$\mathfrak{I}_z(l) = \int_0\limits^{l}\dd\ell\left(1-\frac{(N^2_c-1)\epsilon}{2N_c N_f}w(\ell)\right) = l-\frac{3(N^2_c-1)^\frac{1}{3}}{2^\frac{14}{3}(h^{*}_5)^\frac{1}{3}}l^\frac{1}{3}$,
$\mathfrak{I}_{\psi}(l) = -\mathfrak{I}_z(l) 
+\frac{16\sqrt{2}\sqrt{h^{*}_5}(N^2_c-1)\epsilon^\frac{3}{2}}{N^\frac{3}{2}_cN^\frac{3}{2}_f}\int_0\limits^{l}
\dd\ell \, w(\ell)^\frac{3}{2} =  -\mathfrak{I}_z(l) +\frac{1}{2}\log(l)$.
Using the fact that the fermion Green's function reduces 
to the bare one in the small $w(l)$ limit,
we obtain the scaling form of the Green's function for $n=1$
in the low energy limit with $e^{\mathfrak{I}_{z} (  \log(1/|\mathbf{K}|)  )}\vec{k}\sim 1$, 
\begin{align}
G_1(\mathbf{K},\vec{k}) = \frac{1}{iF_{\psi}(|\mathbf{K}|)}
\frac{1}{	F_{z}(|\mathbf{K}|)	\boldsymbol{\Gamma}\cdot\mathbf{K}
     +\gamma_{d-1} \left[   \frac{\pi N_c N_f}{4\epsilon(N^2_c-1)}
	\frac{k_x}{\log(1/|\mathbf{K}|)}+k_y \right]   },
\end{align}
where 
\begin{align}
v(l) = w(l)^\frac{3}{2}\sqrt{\frac{\lambda (l)}{x(l)}}\approx  \frac{\pi N_c N_f}{4(N^2_c-1)}\frac{1}{\epsilon }\frac{1}{l}
\end{align}
and $F_{z}(|\mathbf{K}|)$, $F_{\psi}(|\mathbf{K}|)$ are given by Eqs. (\ref{eq:Fz}) and (\ref{eq:Fpsi}), respectively.

The Green's function for the boson satisfies the renormalization group equation\cite{PhysRevB.91.125136},
\begin{align}\label{eq:EqgBoson}
\left[z\mathbf{Q}\cdot\frac{\partial}{\partial\mathbf{Q}}+\vec{q}\cdot\frac{\partial}{\partial \vec{q}} - \beta_w\frac{\partial}{\partial w} - \beta_x\frac{\partial}{\partial x} - \beta_\lambda\frac{\partial}{\partial \lambda}-(2\eta_\phi+z(d-1)-(d+1))\right]D(q;w,x,\lambda)=0,
\end{align}
which is solved by
\begin{align}\label{eq:SolD}
D(q;w_0,x_0,\lambda_0) &= \exp\left(-\mathfrak{I}_\phi(l) \right)D(e^{l}\mathbf{Q},e^{\mathfrak{I}_z(l)}\vec{q};w(l),x(l),\lambda(l)).
\end{align}
Here $\mathfrak{I}_z(l)$ is defined in Eq. (\ref{eq:IZ}) and 
\begin{align}
\mathfrak{I}_{\phi}(l) &= \int_0\limits^{l}\dd\ell \left(\frac{2\eta_\phi(\ell)+z(\ell)(2-\epsilon)-(4-\epsilon)}{z(\ell)}\right)\label{eq:IPhi},
\end{align}
with 
\begin{align}
\eta_{\phi}&=\frac{\epsilon}{2}+\frac{((N^2_c-1)(\epsilon-4) + 4)\epsilon}{4N_c N_f} w + \frac{16\sqrt{2}\sqrt{h^{*}_5}(N^2_c-1)(4-\epsilon)\epsilon^\frac{3}{2}}{N^\frac{3}{2}_cN^\frac{3}{2}_f} w^\frac{3}{2}.
\end{align}
From 
$\mathfrak{I}_{\phi}(l) = -2\mathfrak{I}_z(l)
+\int\limits^{l}_{0}\dd\ell\left(\epsilon-\frac{(N^2_c-3)\epsilon}{N_c N_f}w(l)\right)
=-2\mathfrak{I}_z(l)
+\epsilon l-\frac{3(N^2_c-3)}{2^\frac{11}{3}(h^{*}_5)^\frac{1}{3} (N^2_c-1)^\frac{2}{3}}l^\frac{1}{3},
$
the scaling form of the boson propagator is obtained to be
\begin{align}
D(q;w_0,x_0,\lambda_0)&=\exp\left(2\mathfrak{I}_z(l) -\epsilon l +\frac{3(N^2_c-3)}{2^\frac{11}{3}(h^{*}_5)^\frac{1}{3} (N^2_c-1)^\frac{2}{3}}l^\frac{1}{3}\right)D(e^{l}\mathbf{Q},e^{\mathfrak{I}_z(l)}\vec{q};w(l),x(l),\lambda(l)),
\end{align}
where $l=\log(1/|\mathbf{Q}|)$ with $e^{\mathfrak{I}_z(l)}\vec{q}\sim 1$.


\bibliographystyle{apsrev4-1}
\bibliography{sdw}

\end{document}